%% file: Sample.tex
\documentclass{CUP-JNL-DTM}%

\usepackage{graphicx}
\usepackage{multicol,multirow}
\usepackage{amsmath,amssymb,amsfonts}
\usepackage{mathrsfs}
\usepackage{amsthm}
\usepackage{rotating}
\usepackage{appendix}
\usepackage{tikz}
\usetikzlibrary{positioning}
\usetikzlibrary{shapes.geometric, arrows}
\usepackage{xcolor}
\usepackage[numbers]{natbib}
\setcitestyle{round}
\usepackage{ifpdf}
\usepackage[T1]{fontenc}
\usepackage{newtxtext}
\usepackage{newtxmath}
\usepackage{textcomp}
\usepackage{xcolor}
\usepackage{lipsum}
\usepackage[colorlinks,allcolors=blue]{hyperref}
\usepackage[linesnumbered,ruled,vlined]{algorithm2e}

\theoremstyle{definition}

\numberwithin{equation}{section}

\jname{Data-Centric Engineering}
\articletype{RESEARCH ARTICLE}
\jyear{2025}

\begin{document}

\begin{Frontmatter}

\title[Article Title]{Mixed data-source transfer learning for a turbulence model augmented physics-informed neural network}

\author[1]{Christian Toma}
\author[1]{Bharathram Ganapathisubramani}
\author[1]{Sean Symon}

\authormark{Toma \textit{et al}.}

\address[1]{\orgdiv{Department of Aeronautics and Astronautics}, \orgname{University of Southampton}, \orgaddress{\city{Southampton}, \postcode{SO16 7QF},  \country{United Kingdom}}}

\authormark{Toma et al.}



\abstract{Physics-informed neural networks (PINNs) have recently emerged as a promising alternative for extracting unknown quantities from experimental data. Despite this potential, much of the recent literature has relied on sparse, high-fidelity data from direct numerical simulations (DNS) rather than experimental sources like particle image velocimetry (PIV), \textcolor{black}{which are not suitable for validating all reconstructed quantities. In the case of PIV, for example, pressure is not directly measured and the data have imperfections such as noise contamination or a limited field of view.} \textcolor{black}{To overcome these limitations, we present a novel methodology where PINNs are first trained on a RANS simulation such that it learns all states at every location in the domain. We then apply transfer learning which updates the PINN using sub-sampled PIV data. The resulting predictions are in significantly better agreement with the full PIV dataset than PINNs which are trained on experimental data only.} This work builds on the recent literature by integrating a Spalart-Allmaras turbulence model and applying hard constraints to the no-slip wall boundary condition. We apply this new methodology to a two-dimensional NACA 0012 airfoil inclined at an angle of attack, $\alpha$ = 15\textdegree, for two Reynolds numbers of Re = 10,000 and 75,000. \textcolor{black}{The proposed methodology is initially validated using large eddy simulation (LES) data and then demonstrated on experimental PIV data.} Our transfer learning approach results in improved predictions and a reduction in training time when compared to using a random network initialisation.}
\end{Frontmatter}

\section*{Impact Statement}
In fluid dynamics, data assimilation can bridge the gap between experiments and high-fidelity simulations, fixing some of the shortcomings of both. This work extends the recent work using physics-informed neural networks (PINNs) to experimental datasets of higher Reynolds number stalled airfoil flows. Using our proposed methodology, we have shown the ability of PINNs to reconstruct the entire flow field from sparse measurements and predict key surface quantities, such as pressure, without providing reference data.


\input{1-introduction}

\input{2-methodology-v3}
\input{3-setup-v4}

\input{4-results}
\input{5-conclusion}

\begin{appendix}
\input{appendix-a}
\end{appendix}

\begin{Backmatter}

\paragraph{Acknowledgments}
We are grateful for the technical assistance of Uttam Cadambi Padmanaban for producing the LES data and the University of Southampton's High Performance Computing Facility, IRIDIS X, for the technical support and computational resources used in the completion of this work.

\paragraph{Funding Statement}
We gratefully acknowledge funding from the School of Engineering at the University of Southampton for CT's PhD studentship.

\paragraph{Competing Interests}
The authors declare no conflict of interest.

\paragraph{Data Availability Statement}
Pertinent data will be made available with a DOI upon publication at: \url{https://doi.org/10.5258/SOTON/D3539}. 

\paragraph{Ethical Standards}
The research meets all ethical guidelines, including adherence to the legal requirements of the study country.

\paragraph{Author Contributions}
CT contributed to the implementation of the PINN codes, creation of the associated training procedures, analysing the results and writing several drafts. BG and SS were responsible for conceptualisation, funding acquisition, editing of drafts and project management.

\paragraph{Use of artificial intelligence (AI) tools}
Artificial intelligence tools (ChatGPT) were used only for writing refinement in this research.


\end{Backmatter}

\end{document}

%% file: 1-introduction.tex
\section{Introduction} \label{ch:intro}
Computational fluid dynamics (CFD) simulations encapsulate a range of modelling techniques ranging from low-cost Reynolds-averaged Navier-Stokes (RANS) to state-of-the-art direct numerical simulations (DNS) that take vast computational resources to solve (Doran \citeyear{Doran2013}). The weakness of cheaper simulations is that they rely on modelling assumptions to balance the computational cost and fidelity of the data obtained. Alternatively, experimental data has the potential to provide the ground truth from direct measurements of the flow. However, noise and corruption degrade the experimental data and provide varying uncertainty regarding its reliability. Consequently, neither simulations nor experiments alone can provide complete information about complex flows, and this inherent gap has motivated the development of alternative methodologies to address these challenges.

More recently, a third path proposes to bridge the simulated and experimental worlds using optimisation-based approaches to fuse them together and incorporate their respective advantages. One of the latest approaches to solving this problem is physics-informed neural networks (PINNs). Initially proposed by Raissi \textit{et al.} \citeyearpar{Raissi2019}, PINNs utilise modern advancements in machine learning and neural networks to directly solve partial differential equations (PDEs) by leveraging automatic differentiation to compute the necessary gradients. \textcolor{black}{At its core, the assumption is that a neural network can approximate a solution to a non-linear function, in this case, the Navier-Stokes equations} (Hornik \textit{et al.,} \citeyear{Hornik1989}). The differentiating factor between PINNs and CFD is that PINNs directly compare predictions to reference data at discrete locations to find a solution for the entire field.

One of the foundational papers on the PINN methodology and the application of sparse reference data to reconstruct the mean flow quantities is by Sliwinski and Rigas \citeyearpar{Sliwinski2023}. In this work, they investigated predicting a \textcolor{black}{time-averaged} cylinder flow at Reynolds number of Re = 150 from a sparse sampling of a DNS solution. To do so, they used the forcing representation of the Reynolds stresses initially proposed by Foures \textit{et al.} \citeyearpar{Foures2014} to reduce the discrepancy between unknowns and equations to solve. By decomposing the forcing via the Helmholtz decomposition, the disparity of unknowns to equations is one. Even with an underdetermined system of equations, this PINN formulation yielded improved predictions of the mean velocity field with uniformly sparse data of two points per cylinder diameter. Furthermore, the results highlighted the ability of PINNs to work with sparse and noisy data. However, the authors noted in the conclusions that there were discrepancies at the surface of the cylinder. 

Another approach to simplifying the RANS equations and the Reynolds stress tensor is to apply the Boussinesq approximation that relates the mean velocity gradients to the Reynolds stresses through an added viscosity term, referred to as the eddy viscosity, $\nu_t$ (Boussinesq \citeyear{Boussinesq1877}). von Saldern \textit{et al.} \citeyearpar{vonSaldern2022} used this approach to assimilate a swirling turbulent jet flow from a reference \textcolor{black}{stereo-PIV} dataset. The network implicitly derived an eddy viscosity model from the data, which, when used to calculate the Reynolds stresses, had good agreement with the data.


Rather than allow the PINN to indirectly construct a model for computing the Reynolds stresses, an alternative approach is to incorporate an established turbulence model, as is common in other numerical methods. Patel \textit{et al.} \citeyearpar{Patel2024} extended the equations to include the Spalart–Allmaras turbulence model, in which an additional variable, $\tilde{\nu}$, is solved. This variable represents a modified form of the turbulent kinematic viscosity and is used to compute the eddy viscosity term, $\nu_t$, which accounts for the additional momentum transfer induced by turbulent fluctuations. This formulation does not change the disparity of unknowns to equations but provides further constraints on the PINN to model the physics accurately. In this work, they applied this new formulation to a periodic hill case for a Reynolds number of Re =  5600. When compared to a baseline PINN without a turbulence model, the addition of a turbulence model improved flow predictions throughout the domain, particularly in the near-wall region and at the separation point (Patel \textit{et al.,} \citeyear{Patel2024}). Others have also implemented a turbulence model into the PINNs equation formulation. Pioch \textit{et al.} \citeyearpar{Pioch2023} applied a k-$\omega$, as well as other mixing length-based models, to a backwards-facing step case at a Reynolds number of Re = 5100 and found good agreement with reference data but noted the difficulty in training stability, resulting in unphysical predictions. 

In \textcolor{black}{many} of these cases, the reference data used to train the PINN originated from a DNS solution as opposed to experimental measurements. Even with the current state-of-the-art DNS solvers, the computational cost for generating this data is prohibitively expensive, particularly for parametric studies or rapid design iterations (Pope \citeyear{Pope2001}). \textcolor{black}{As such, experimental data might be more appropriate despite its limitations.} Some studies have explored replicating the noise present in experimental data, but relatively few have directly integrated raw experimental data into PINNs. One \textcolor{black}{example} by Steinfurth \textit{et al.} \citeyearpar{Steinfurth2024} trained a PINN model using \textcolor{black}{planar PIV collected at multiple planes}, surface pressure and wall shear-stress measurements from a backwards-facing \textcolor{black}{ramp} model. \textcolor{black}{These measurements were aggregated and used to train a three-dimensional PINN with a mixing length model to approximate the Reynolds shear stresses instead of one of the aforementioned eddy viscosity models.} When training a PINN, the loss function incorporates evaluations at both data points from a reference data set and randomly generated collocation points. In the context of PINNs, collocation points are equivalent to cells in a CFD solver, which provide locations to assess how closely the predicted quantities fit the governing equations and boundary conditions. In Steinfurth \textit{et al.} \citeyearpar{Steinfurth2024}, with the amount of data points in the domain, they required much fewer additional collocation points to evaluate the physics loss and boundary conditions. Furthermore, the authors found that the predictions improved overall by significantly suppressing the contribution of the physics loss to the total loss. \textcolor{black}{This approach lends itself to a data-rich assimilation problem and was shown to be effective, in particular, for fixing stitching errors in the reference PIV dataset.} \textcolor{black}{Other attempts to integrate a turbulence model within a PINN include the work by Villi{\'e} \textit{et al.} \citeyearpar{Villie2025}. Here, the authors applied a Spalart-Allmaras turbulence model to enhance experimental MRI measurements of a transitioning stenosis flow. The PINN reduced errors associated with the measurement technique in the velocity components and found reasonable agreement with the simulations for other quantities, such as pressure and Reynolds stresses.}

\textcolor{black}{Klopsch \textit{et al.} \citeyearpar{Klopsch2025} expanded the work of Patel \textit{et al.} \citeyearpar{Patel2024} to apply PINNs with the same formulation of the Spalart-Allmaras equations but to experimental PIV measurements of a flow past a Boeing Gaussian Bump at two Reynolds numbers of Re = $1 \times 10^6$ and $2 \times 10^6$, respectively. These two Reynolds numbers represent a mostly attached and a fully separated flow, respectively. The authors provided windows of time-averaged PIV data at different streamwise locations along the Gaussian bump. The PINN then reconstructed the continuous velocity field across the domain, alongside predicting the pressure and eddy viscosity fields. Whilst the obtained fields were physically consistent, it was noted that the assimilated eddy viscosity did not fully conform to the Spalart-Allmaras model.}

For most existing PINN methodologies, the network weights and biases are initiated randomly and sampled from a distribution based on the activation function used. In the broader machine learning community, it is common to start from a pre-trained model developed to solve a similar task, for example, in computer vision or text-based tasks (Too \textit{et al.,} \citeyear{Too2019}; Ziegler \textit{et al.,} \citeyear{Ziegler2019}). This process, known as transfer learning, involves re-training or fine-tuning a model with a new, domain-specific dataset. The benefit of this is reducing computational cost by reducing the subsequent training time and often improved results over building a model directly (Torrey and Shavlik \citeyear{Torrey2010}; Weiss \textit{et al.,} \citeyear{Weiss2016}). 

\textcolor{black}{In this work, we apply transfer learning to PINNs such that they reconstruct mean flows from limited experimental measurements. The PINN enforces the incompressible RANS equations with a Spalart-Allmaras turbulence model. It is first trained on a baseline RANS simulation such that it successfully predicts all states including the eddy viscosity and pressure. The PINN is subsequently transferred to sub-sampled LES and PIV data to reconstruct all states inside and outside the experimental domain. The rest of the paper is organised as follows.} In Section~\ref{ch:methods}, we detail the PINN formulation and specifics of the configuration for this work. Section~\ref{ch:setup} outlines the new training methodologies as well as the datasets used throughout this work. Section~\ref{ch:results} showcases the results and discusses some of the fundamental challenges when using PINNs in this application. \textcolor{black}{Section~\ref{ch:conclusion} summarises the findings from applying these new methodologies and discusses some potential extensions.}

%% file: 2-methodology-v3.tex
\section{PINN and problem setup} \label{ch:methods}
Like many other data assimilation techniques, the PINN framework fits sparse, high-fidelity measurements to a set of governing equations (Raissi \textit{et al.,} \citeyear{Raissi2019}). Given the scope of this work is to assimilate flow measurements in 2D, we construct the physics loss of the PINN from the 2D RANS equations. The methodology presented here builds upon recent work by Sliwinski and Rigas \citeyearpar{Sliwinski2023} and Patel \textit{et al.} \citeyearpar{Patel2024}. Section~\ref{sec:eqs} focuses on the governing equations used in this work, with Section~\ref{sec:loss-form} describing how these equations integrate within the loss function. Section~\ref{sec:hard-bcs} outlines the extension to boundary conditions through the application of hard constraints. Section~\ref{sec:net-arch} covers the details of the network architecture, and Section~\ref{sec:domain} aggregates \textcolor{black}{the} various aspects of the PINN and provides specific information about the implementation for the work presented through the result sections. 

\subsection{Governing equations} \label{sec:eqs}
For a mean flow problem, we can apply the Reynolds decomposition to split the flow quantities into their respective steady ($\overline{U_i}, \overline{P}$) and unsteady ($u'_i, p'$) parts. Once substituted into the Navier-Stokes equations and averaged in time, we obtain the RANS equations

\begin{align}
     \frac{\partial \overline{U_i}}{\partial x_i} = 0,
     \nonumber
     \\
     \overline{U_j}\frac{\partial \overline{U_i}}{\partial x_j}
     + \frac{1}{\rho} \frac{\partial \overline{P}}{\partial x_i}
     - \frac{\partial (2 \nu S_{ij}) }{\partial x_j}
     + \frac{\partial \overline{u'_i u'_j}}{\partial x_j} = 0,
    \label{eq:baseline-rans}
\end{align}

\noindent where $\overline{U_i}$ is the mean velocity, $P$ is pressure, $\rho$ is density and $\nu$ is kinematic viscosity. The two remaining terms represent the strain rate tensor, $S_{ij} = \frac{1}{2}\left( \frac{\partial \overline{U_i}}{\partial x_j} + \frac{\partial \overline{U_j}}{\partial x_i}\right)$, and the divergence of the Reynolds stress tensor $\overline{u'_iu'_j}$. For a 2D problem, Equation~\ref{eq:baseline-rans} provides three equations with six unknowns, creating an under-determined system to solve. This issue is known as the closure problem. Within the turbulence modelling and data assimilation communities, numerous approaches, with varying levels of complexity, have been proposed as solutions to the closure problem. In the work presented here, we will consider two approaches: one model-free and the other modelled using a Spalart-Allmaras turbulence model.

For the model-free approach, we allow the PINN to determine the Reynolds stresses through the training process. In this case, the output layer contains a node for each component of the Reynolds stress tensor. However, it is more effective to reduce the disparity between equations and unknowns (Sliwinski and Rigas \citeyear{Sliwinski2023}). Initially proposed by Foures \textit{et al.} \citeyearpar{Foures2014}, the divergence of the Reynolds stresses can be represented as a forcing term,

\begin{equation} \label{eq:helmholtz}
    f_i \equiv -\frac{\partial \overline{u'_i u'_j}}{\partial x_j}
    = \frac{1}{\rho} \frac{\partial \phi}{\partial x_i} + f_{s, i}.
\end{equation}

\noindent Applying the Helmholtz decomposition to $f_i$ results in two terms: a potential, $\phi$ and solenoidal, $f_{s, i}$, component. By combining Equation~\ref{eq:baseline-rans} with Equation~\ref{eq:helmholtz}, we can derive the RANS with forcing (RANS-F) equations, given by

\begin{align}
     \frac{\partial \overline{U_i}}{\partial x_i} = 0,
     \nonumber
     \\
     \overline{U_j}\frac{\partial \overline{U_i}}{\partial x_j}
     + \frac{1}{\rho} \frac{\partial (\overline{P} - \phi)}{\partial x_i}
     - \frac{\partial (2 \nu S_{ij}) }{\partial x_j}
     - f_{s, i} = 0, \nonumber
     \\
     \frac{\partial f_{s,i}}{\partial x_i} = 0.
    \label{eq:rans-f}
\end{align}

As the forcing term, $f_{s, i}$, is solenoidal, we can also specify that the forcing must be divergence-free. Additionally, to reduce the number of unknowns, we can combine the pressure with $\phi$ as a single term ($\overline{P} - \phi$). The resulting system remains under-determined, but the number of unknowns has been reduced to five with four equations. The reduction in unknowns, whilst attempting to mitigate one problem, creates the issue that the pressure remains coupled with $\phi$. Therefore, even with a fully known velocity field and subsequently making the system determined, the predicted pressure may not be equivalent to the true pressure as the influence of $\phi$ is unknown (Foures \textit{et al.,} \citeyear{Foures2014}; Sliwinski and Rigas \citeyear{Sliwinski2023}).

The second approach to solving the closure problem is empirically modelling the Reynolds stresses through a turbulence model. This approach is commonly seen in numerical solvers and relies on the Boussinesq approximation to relate the Reynolds stresses to a combination of the mean velocity gradients and a modelled eddy viscosity term, $\nu_t$, as given by

\begin{equation} \label{eq:boussinesq}
    {\color{black}\overline{u'_i u'_j} = -2\nu_t S_{ij} +\frac{2}{3}k\delta_{ij}},
\end{equation}

\noindent \textcolor{black}{for an incompressible flow,} where $k$ is the turbulent kinetic energy and $\delta_{ij}$ is the Kronecker delta (Boussinesq \citeyear{Boussinesq1877}). In this work, following recent literature, a Spalart-Allmaras (SA) turbulence model was selected to model the eddy viscosity term (Franceschini \textit{et al.,} \citeyear{Franceschini2020}; Patel \textit{et al.} \citeyear{Patel2024}). For a mean flow problem, the transport equation simplifies to,

\begin{equation} \label{eq:sa-transport}
    \overline{U_j}\frac{\partial \tilde{\nu}}{\partial x_j}
    - c_{b1}\tilde{S}\tilde{\nu} + c_{w1}f_{w} \left( \frac{\tilde{\nu}}{d} \right)^2
    - \frac{1}{\sigma} \left[\frac{\partial}{\partial x_j} \left((\nu + \tilde{\nu}) \frac{\partial \tilde{\nu}}{\partial x_j} \right) + c_{b2}\left(\frac{\partial \tilde{\nu}}{\partial x_j} \right)^2 \right] = 0.
\end{equation}

This formulation provides one additional unknown, $\tilde{\nu}$, and one additional equation (Spalart and Allmaras \citeyear{Spalart1994}). For the Spalart-Allmaras transport equation, Equation~\ref{eq:sa-transport}, the mean velocity gradients predominantly drive the production and destruction of eddy viscosity through the domain. \textcolor{black}{Therefore, for the PINN, the sensitivity to the mean velocity gradients is increased to solve this equation correctly.} \textcolor{black}{The following equation relates this additional term to the eddy viscosity in the momentum equations}

\begin{subequations}
\label{eq:nutile-nut}
\begin{align}
    {\color{black} \nu_t = \tilde{\nu} f_{v1}}, \label{eq:nutile-nut-a} \\
    {\color{black} f_{v1} = \frac{\chi^3}{\chi^3 + c_{v1}^3}}, \label{eq:nutile-nut-b} \\
    {\color{black} \chi = \frac{\tilde{\nu}}{\nu}}, \label{eq:nutile-nut-c}
\end{align}
\end{subequations}

\noindent \textcolor{black}{where $c_{v1} = 7.3$.} With this approach, the forcing, $f_i$, from Equation~\ref{eq:helmholtz}, will still represent the divergence of the Reynolds stresses. The key difference is now the forcing term is a corrective forcing, $f_{c, i}$, which can be decomposed in the same method to its respective solenoidal and potential components, as shown

\begin{equation} \label{eq:eddy-forcing}
    f_{i} = \frac{\partial(2 \nu_t S_{ij})}{\partial x_j} + f_{c, i} = \frac{\partial(2 \nu_t S_{ij})}{\partial x_j} + \frac{1}{\rho} \frac{\partial \phi}{\partial x_i} + f_{s, i}.
\end{equation}

This corrective forcing formulation aims to capture any differences between the turbulence model and any measured data. Notably, the solenoidal forcing, $f_{s, i}$ from Equation~\ref{eq:helmholtz} will not be the same, but $f_i$ should be when converged. As the purpose of the turbulence model is to compute the Reynolds stresses, the forcing, $f_{s, i}$, in Equation~\ref{eq:eddy-forcing} captures any components that the model cannot represent due to the constraints of the Boussinesq approximation. \textcolor{black}{Whilst this does not resolve the problem of an under-determined system of equations to solve, the benefit of a turbulence model is to aid in improving predictions in higher Reynolds number flows as would be done in traditional numerical methods.} Previously, the equivalent forcing term in Equation~\ref{eq:helmholtz} would need to do all of this. To promote the PINN's use of the turbulence model, we penalise the production of forcing by minimising the magnitude of the sum of all solenoidal forcing terms. Once substituted into the momentum equation in Equation~\ref{eq:baseline-rans}, the final definition for the RANS-SA equations is 

\begin{equation} \label{eq:sa-momentum} 
    \overline{U_j}\frac{\partial \overline{U_i}}{\partial x_j}
    + \frac{1}{\rho} \frac{\partial (\overline{P} - \phi)}{\partial x_i}
    - \frac{\partial (2 (\nu + \nu_t) S_{ij}) }{\partial x_j}
    - f_{s, i} = 0. 
\end{equation}

\subsection{Loss formulation} \label{sec:loss-form}
Traditionally, deep neural networks solve regression problems by minimising a pointwise comparison between the predicted and reference values through the mean-squared error, commonly referred to as supervised learning. \textcolor{black}{PINNs expand on this methodology by expanding this loss function to include additional supervised and unsupervised terms.} These terms fall into three categories: physics loss, boundary loss and data loss,

\begin{equation} \label{eq:loss_fn}
    \mathcal{L}_{total} = \mathcal{L}_{P} + \mathcal{L}_{B} + \mathcal{L}_{D}.
\end{equation}

The contribution of each component, $\mathcal{L}_{total}$, is then used to evaluate the predicted quantities and update the network's parameters. For the physics loss, $\mathcal{L}_{P}$, we can use the predicted quantities from the network to solve the governing partial differential equations (PDEs). The fundamental assumption for all PINN problems is that the gradients computed from the network \textcolor{black}{with automatic differentiation correctly represent the derivatives of the solution and can be used to enforce the governing PDEs.} For example, by moving all terms of the RANS-F equations to one side, as shown in Equation~\ref{eq:rans-f}, the PINN computed values of each left-hand side become a residual we are minimising. This type of training is unsupervised, as we do not provide reference values to solve this part of the loss function. In practice, we generate random collocation points, analogous to cells of a mesh, that are passed through the network to compute the localised values and their associated gradients. We can express the physics loss as,

\begin{equation} \label{eq:phy-loss}
    \mathcal{L}_{P} = \frac{\lambda_P}{N_P} \sum_{i=1}^{N_P} ||f(\boldsymbol{X}_i)||_{2}^{2},
\end{equation}

\noindent where $N_p$ is the number of collocation points evaluated by \textcolor{black}{each equation, $f$,} and its overall contribution is weighted by $\lambda_P$. 

For many non-linear PDE problems, a trivial solution is valid, and if left unsupervised, a neural network will identify it as the most optimal solution. To prevent this, we must apply additional constraints to the problem through supervised methods. One approach is to provide boundary conditions with reference values that are physically derived, such as a no-slip wall. For the PINN, this acts as a supervised component of the loss function and is evaluated through the MSE as described by,

\begin{equation} \label{eq:bc-loss}
    \mathcal{L}_{B} = \frac{\lambda_B}{N_{B,i}} \sum_{i=1}^R \left[\sum_{j=1}^{N_{B,i}} \left(\boldsymbol{X}_{i, j} - \boldsymbol{\hat{X}}_{i, j}\right)^2\right],
\end{equation}

\noindent where $\lambda_B$ is the weighting of the boundary loss to the total loss. The important distinction here is the source of $N_B$. These points are from generated collocation points, like the physics loss, specifically those on a defined boundary. The specific boundary conditions, construction of the domain and the generation of the collocation points are discussed in detail in Section~\ref{sec:domain}.

Like the boundary loss, the data loss provides reference data at discrete locations but from a measured data set rather than our physical understanding of the flow. The sample data can be sourced from either numerical or experimental datasets and is evaluated through the MSE as described,

\begin{equation} \label{eq:data-loss}
    \mathcal{L}_{D} = \frac{\lambda_D}{N_{D,i}} \sum_{i=1}^R \left[\sum_{j=1}^{N_{D,i}} \left(\boldsymbol{X}_{i, j} - \boldsymbol{\hat{X}}_{i, j}\right)^2\right],
\end{equation}

\noindent where $i$ represents the flow variable in the reference data, $\boldsymbol{X}$, and the prediction, $\boldsymbol{\hat{X}}$, at each $j$-th data point. The contribution of $\mathcal{L}_{D}$ is weighted by $\lambda_D$ and evaluated at $N_D$ data points for each quantity provided. Importantly, the associated location of this data point, used as an input to the network, does not necessarily need to be in the collocation points. In this case, a secondary forward pass computes the necessary predicted values for the loss function, and the network aggregates all of the aforementioned loss components for backpropagation.

\subsection{Hard constraints} \label{sec:hard-bcs}
In Section~\ref{sec:loss-form}, each component of the total loss function is implicitly a soft constraint that is minimised towards zero. Therefore, the optimiser will update the network parameters by minimising the components that contribute most to the total loss, either through its residual or by that residual being amplified by its associated multiplier, $\lambda$. These constraints are referred to as \textit{soft} constraints, as the predicted quantity may only satisfy the constraint to a converged residual. This behaviour may be preferable for certain constraints, such as data loss, as the data may contain noise or outliers. However, not satisfying these constraints can produce unphysical or undesired results for certain boundary conditions. A practical example of this would be the difference in lift generated for a wing in ground effect with a slip or no-slip wall.

To overcome potentially unphysical behaviour, \textit{hard} constraints are employed along these boundaries to ensure the predicted values match those prescribed by the boundary condition. Unlike many other data assimilation methods, the practical implementation for neural networks is not trivial due to its stochastic nature. One proposed method uses an augmented Lagrangian method (ALM) (Lu \textit{et al.,} \citeyear{Lu2021}). Fundamentally, these work similarly to the loss multipliers; however, there are various methodologies for updating the value of the Lagrange multiplier to enforce the constraint more firmly during training.

In the work presented here, we utilise the approach proposed by Basir and Senocak \citeyearpar{Basir2023}. Here, the authors updated the Lagrange multipliers using the RMSprop algorithm (Hinton \textit{et al.,} \citeyear{Hinton2012}). This unique multiplier per boundary point approach enables a more balanced learning process for complex PDE problems with many constraints without needing significant hyperparameter tuning once applied. However, over a more simplistic approach, such as a monotonically increasing multiplier, this adaptive ALM method increases computational cost in both required training time and memory usage. The increased computational cost comes from the RMSprop algorithm requiring two additional parameters, the momentum and gradient, that update the Lagrange multiplier. The additional parameters also increase computational time, with each parameter having an associated equation to update them per iteration. Additionally, as the multiplier is unique to the collocation point, the number of additional parameters to keep in memory is three times the number of boundary points in a given hard constraint. Therefore, when creating the PINN, it is critical to be selective about which constraints are hard constraints.

\subsection{Network architecture} \label{sec:net-arch}
One of the objectives of this work is to investigate whether the trends captured by one PINN model are generalisable to a new one through transfer learning. Section~\ref{sec:train-tl} details the specifics for applying transfer learning to PINNs, but fundamentally, to enable this, the same network architecture must be used to directly copy all weights and biases from one network to another. Therefore, following recent literature, a fully connected neural network (FCNN) is constructed with two input nodes ($x \ \& \ y$), seven hidden layers with 50 nodes each and an output layer of six nodes with a hyperbolic tangent (tanh) activation function on all but the output layer. These six nodes relate to each of the unknown quantities in the 2D RANS-SA equation formulation ($\overline{U}, \overline{V}, (P-\phi), f_{s,x}, f_{s,y} \ \& \ \sqrt{\tilde{\nu}}$). There are two notable points regarding this formulation. Firstly, as in Patel \textit{et al.} \citeyearpar{Patel2024}, to ensure $\tilde{\nu}$ is positive, the output from the network is squared, hence the need to predict $\sqrt{\tilde{\nu}}$. Secondly, to enable transfer learning between equations, for models trained with the RANS-F equation set, $\tilde{\nu}$ is still included as a node on the output layer. Any associated boundary conditions and data loss terms for $\tilde{\nu}$ are still applied to softly limit the range of predicted values, but the physics loss does not evaluate the field across the domain.

\subsection{PINN configuration} \label{sec:domain}
\begin{figure}[h!]
    \centering
    \includegraphics[width=\textwidth]{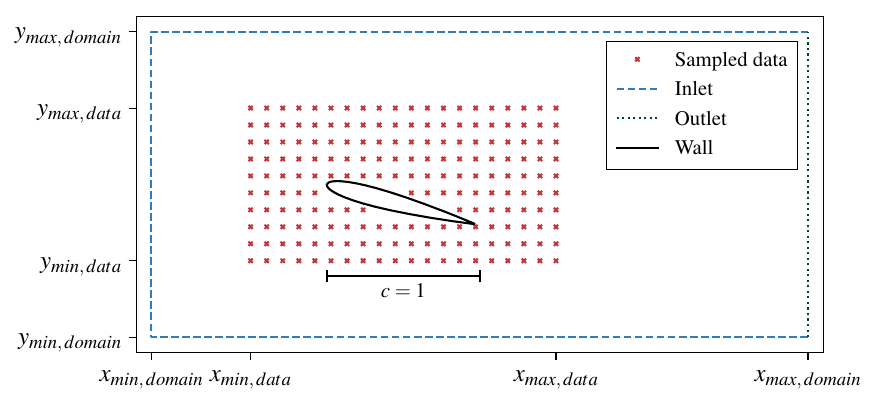}
    \caption{Schematic of domain with representative data loss points within inner data "box" and boundaries for applied boundary conditions}
    \label{fig:annot-domain}
\end{figure}

Like any numerical method, the domain and boundary conditions applied are key to obtaining physically meaningful results. As mentioned in Section~\ref{sec:loss-form}, the purpose of the domain is to provide collocation points to evaluate the network across each loss component. \textcolor{black}{A total of 180,000 training points (150,000 domain and 30,000 boundary points) were used to evaluate the physics and boundary losses, with 25,000 (20,000 domain and 5,000 boundary points) validation points used to evaluate these metrics and the network's performance. All points were randomly generated using a Hammersley distribution, either through the domain or along its edge for boundary points. The number of collocation points was chosen to minimise training time and operate within the GPU's memory constraints.} In the work presented here, we will be investigating the application of PINNs to a stalled airfoil flow across a range of Reynolds numbers and data fidelities. Specifically, a NACA 0012 airfoil at an angle of attack, $\alpha$ = 15\textdegree \ is selected at two Reynolds numbers of Re = 10,000 and 75,000. As shown by the schematic in Figure~\ref{fig:annot-domain}, the domain is constructed from an external box and an internal airfoil. Along the outer edges of the external box, we can prescribe the inlet and outlet conditions and treat the internal edge of the airfoil as a solid wall. This domain geometry is agnostic of any dataset or Reynolds number changes and will be fixed throughout all models. To ensure the PINN obeys the no-slip condition, we apply hard constraints along the surface of the airfoil. Soft constraints are applied to all other boundary conditions, including the data loss, to mitigate the increased computational cost and enable the flexibility of a PINN to handle noisy or erroneous data. Table~\ref{tab:bc-summary} summarises all PINN configurations' applied boundary conditions.

\begin{table}[h!]
    \centering
    \tabcolsep=20pt%
    \TBL{\caption{Boundary conditions applied on all PINN models including type of constraint enforcement \textit{box}\label{tab:bc-summary}}}
    {\begin{fntable}
        \begin{tabular}{lcc}
            \toprule
            \TCH{Boundary} & \TCH{Enforcement} & \TCH{Constraints} \\ \midrule
            \TCH{Inlet}    & Soft & \TCH{$\overline{U} = 1$, $\overline{V} = 0$, $\tilde{\nu} = 4\nu_\infty$} \\
            \TCH{Airfoil (wall)} & Hard & \TCH{$\overline{U} = \overline{V} = f_{s,x} = f_{s,y} = \tilde{\nu} = 0$} \\
            \TCH{Outlet} & \multicolumn{1}{c}{--} & \multicolumn{1}{c}{--} \\ \botrule
        \end{tabular}
    \centering
    \end{fntable}}
\end{table}

\textcolor{black}{For all cases presented here, the domain remained fixed, with a length of 15c (-5c to 10c) and a height of 10c (-5c to 5c), with the airfoil leading edge located at the origin. This study considers four datasets in total, including two RANS simulations (R10 $\&$ R75), and two high-fidelity datasets from LES (L10) and PIV (P75), respectively. These will provide the reference data for the baseline models, as well as the transfer learning models that will be pre-trained on RANS and then fine-tuned using the high-fidelity data. With regards to the PINN configuration and training methodologies for these various models, full details are provided in Section~\ref{ch:setup}. Specifically, Table~\ref{tab:baseline-data} and Table~\ref{tab:tl-data}  describe the dimensions of the data box, the sub-sampling of the reference data, and the quantities used for training as part of the data loss, and those evaluated throughout the domain to validate the physical consistency of the PINN.}


\begin{table}[h!]
\tabcolsep=20pt%
{\color{black}
\TBL{\caption{Summary of loss function weights during Adam and L-BFGS optimisation stages for all soft-enforced constraints\label{tab:lambdas}}}
{\begin{fntable}
\begin{tabular}{llccc}\toprule%
\TCH{Optimiser} & \TCH{Dataset} & \TCH{$\lambda_P$} & \TCH{$\lambda_B$} & \TCH{$\lambda_D$} \\\midrule
\TCH{Adam} & \TCH{R10}& \multirow{2}{*}{1} & \multirow{2}{*}{10} & \multirow{2}{*}{0.01}\\
\TCH{} & \TCH{R75}& & & \\\cmidrule{2-5}
\TCH{} & \TCH{L10}& \multirow{2}{*}{1} & \multirow{2}{*}{10} & \multirow{2}{*}{10} \\
\TCH{} & \TCH{P75}&  &  & \\\cmidrule{2-5}
\TCH{L-BFGS} & \TCH{All} & 1 & 1 & 1 \\
\botrule
\end{tabular}%
\centering 
\end{fntable}}
}
\end{table}

Table~\ref{tab:lambdas} details the base weights for the loss function described in Equation~\ref{eq:loss_fn}. The weights detailed here only apply to soft enforced boundary conditions. For hard constrained boundary conditions, such as the airfoil wall, we assume $\lambda_B = 1$ and the Lagrange multiplier is updated internally per collocation point on the boundary. \textcolor{black}{The data provided, and quantities predicted, are scaled by the freestream velocity, $\overline{U}_{\infty}$, and the airfoil's chord length, c. No additional scaling or normalisation of the inputs or outputs was undertaken here, as seen in other areas of the literature.}

%% file: 3-setup-v4.tex
\section{Transferring to \textcolor{black}{high-fidelity} data} \label{ch:setup}
To apply PINNs to higher Reynolds number problems efficiently using eddy viscosity models, we need to reconsider how we train these models. \textcolor{black}{Section~\ref{sec:ref-data} summarises the datasets used throughout this work for both the baseline and transfer learning models.} Section~\ref{sec:train-baseline} details a new methodology for incorporating the reference RANS eddy viscosity field within the data loss \textcolor{black}{which prevents the PINN from reaching a trivial solution for the eddy viscosity.} Section~\ref{sec:train-tl} discusses how the RANS-trained baseline model can be fine-tuned through transfer learning to high fidelity numerical or experimental datasets. 

\subsection{Reference datasets} \label{sec:ref-data}
To apply the proposed methodologies, we first need to fully define the reference data used for the generation of each model. For each PINN model and its accompanying reference data set, we can group each quantity into driving or driven quantities. Driving quantities are those that appear in the data loss, whereas driven quantities are those that do not feature in the data loss or are derivatives of the PINN's predictions. It is important to make this distinction when analysing the results as we expect driving quantities to be predicted accurately, especially at the data location. However, driven quantities rely on the network to reconstruct them. By comparing the driven quantities for which we have data, we can have greater confidence in the physical interpretability of other quantities for which we do not have data.

\begin{figure}[h!]
    \centering
    \includegraphics[width=\linewidth]{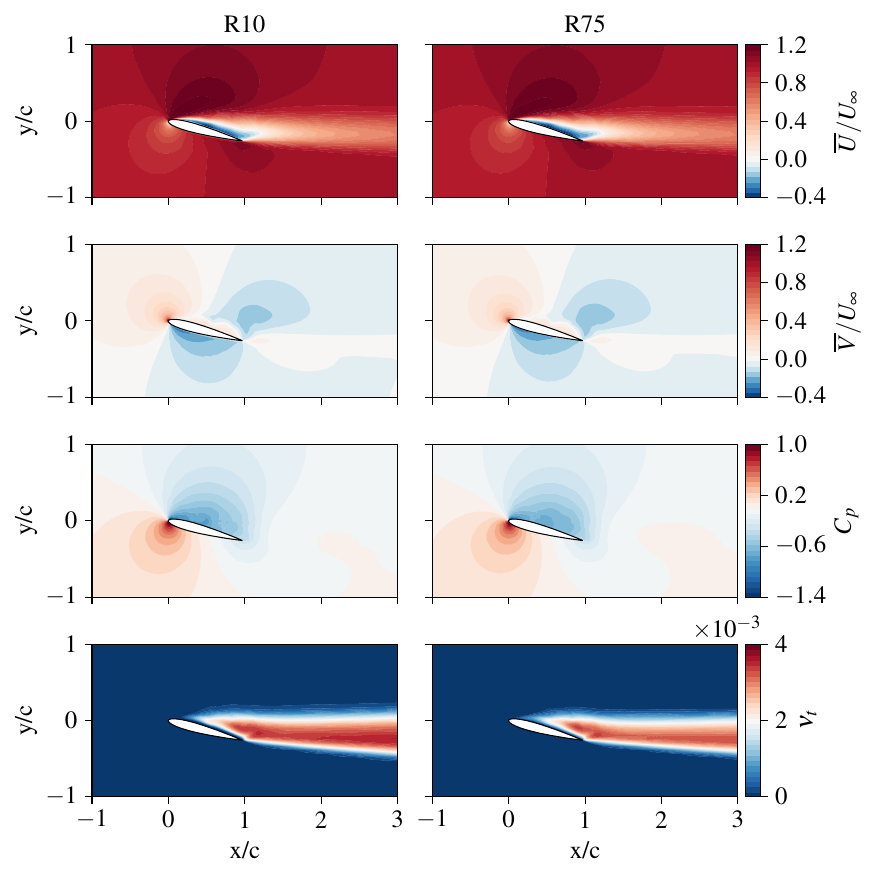}
    \caption{Reference fields for velocity, pressure coefficient, $C_p$, and eddy viscosity, $\nu_t$, fields for a NACA 0012 at $\alpha$ = 15\textdegree \ from RANS at Re = 10,000 (left) $\&$ Re = 75,000 (right)}
    \label{fig:ref-r10-r75-all}
\end{figure}

For the two RANS datasets (R10 and R75), the driving quantities for the baseline models include $\overline{U}, \overline{V}$ and $\tilde{\nu}$ and the driven quantity for comparison is the pressure coefficient, $C_p$\textcolor{black}{, where}

\begin{equation} \label{eq:cp-orig}
    {\color{black}
    C_p = \frac{\overline{P} - \overline{P}_\infty}{q_\infty},}
\end{equation}

\noindent \textcolor{black}{and $q_\infty = \frac{1}{2} \rho \overline{U}^2$. For the PINN, due to the Helmholtz decomposition in Equation~\ref{eq:helmholtz}, the pressure coefficient is calculated by} 

\begin{equation} \label{eq:cp}
    {\color{black}
    C_p = \frac{(\overline{P} - \phi) - (\overline{P} - \phi)_\infty}{q_\infty}.}
\end{equation}

\textcolor{black}{When training with the RANS datasets, no sub-sampling is applied, where all 28,934 cells are used as data points with the driving quantities evaluated at these locations.}

\begin{figure}[h!]
    \centering
    \includegraphics[width=\linewidth]{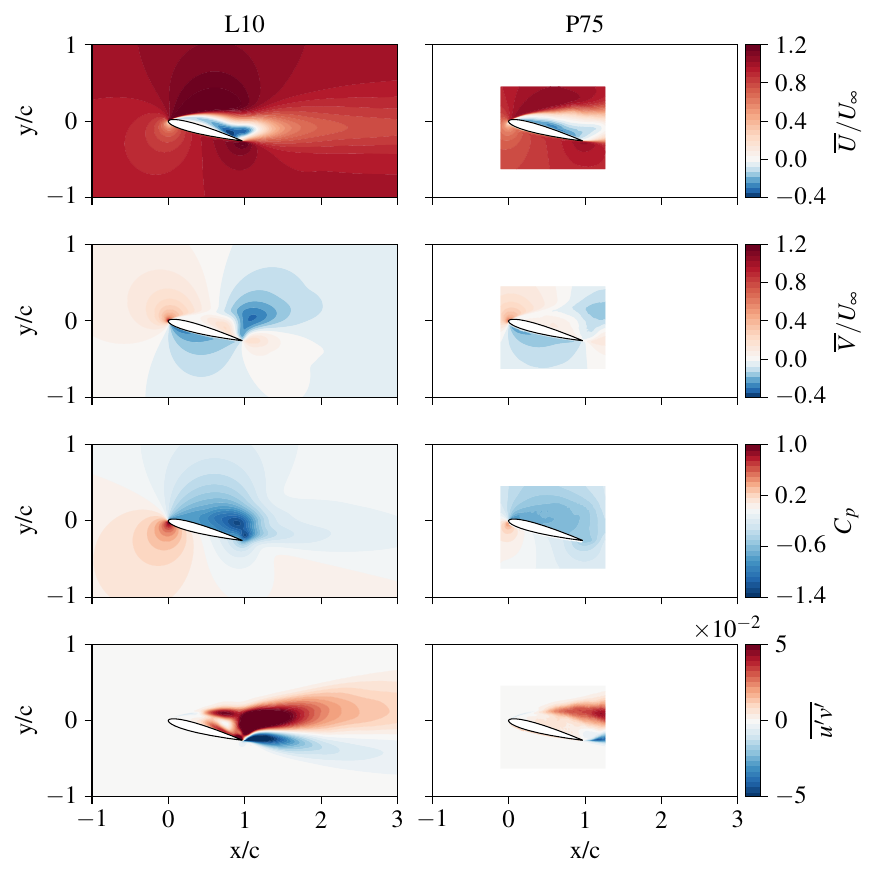}
    \caption{Reference fields for velocity, pressure coefficient, $C_p$, and Reynolds shear stress, $\overline{u'v'}$, fields for a NACA 0012 at $\alpha$ = 15\textdegree \ from LES at Re = 10,000 (left) $\&$ PIV at Re = 75,000 (right). PIV $C_p$ is computed using a Poisson solver and not directly measured}
    \label{fig:ref-l10-p75-all}
\end{figure}

For the two high-fidelity datasets from LES and PIV (L10 and P75, respectively), the driving quantities will only include $\overline{U}$ and $\overline{V}$\textcolor{black}{. The} driven quantities \textcolor{black}{are the} Reynolds shear stress, $\overline{u'v'}$ and $C_p$. The reference data are provided in Figure~\ref{fig:ref-r10-r75-all} and Figure~\ref{fig:ref-l10-p75-all} for comparison both with the baseline models generated using these datasets as well as for transfer learning. \textcolor{black}{When comparing Figure~\ref{fig:ref-r10-r75-all} to Figure~\ref{fig:ref-l10-p75-all}, the necessity for high-fidelity data over RANS becomes apparent. For this stalled airfoil case, RANS does not reliably capture the size of the recirculation bubble and the strength of the reverse flow region.} Table~\ref{tab:baseline-data} summarises the quantities used in the data loss and sampling frequencies through the domain. 

When utilising these \textcolor{black}{high-fidelity datasets, uniform sub-sampling of the data was undertaken to evaluate the PINNs performance with sparse measurements. In this process, there was no interpolation prior to sub-sampling  to ensure that the PINN did not propagate any errors.} \textcolor{black}{The PIV data is from a set of experiments by Carter \textit{et al.} \citeyearpar{Carter2023}. Whilst we sub-sampled the data at 20 points per chord, we dropped any data with clear post-processing errors or near the wall to improve training stability and prediction accuracy. This resulted in a total of 586 data points to evaluate both driving quantities when training with the PIV data. For the LES data, reducing the window size mimics a typical experimental field of view. This was also found to be beneficial for improving predictions in the wake by naturally focusing the optimiser to improve predictions in this region. Given the slightly larger data window, a total of 769 data points were used in training for evaluating the data loss.} 

\begin{figure}[h!]
    \centering
    \includegraphics[width=\linewidth]{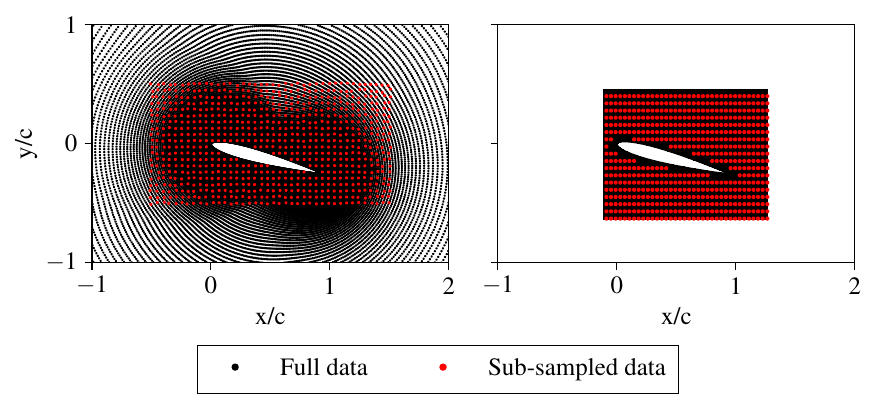}
    \caption{\textcolor{black}{Comparison of original reference data, in black, and sub-sampled data, in red, for the LES (left) and PIV (right) datasets, respectively}}
    \label{fig:sub-sample}
\end{figure}

\textcolor{black}{Figure~\ref{fig:sub-sample} shows the sub-sampled data, in red, against the full available data, in black, for both the LES and PIV datasets. For the LES data, the slight non-uniformity in the sampling is due to selecting the nearest neighbour to a uniform grid of points to avoid any potential interpolation errors.}

\subsection{Baseline training} \label{sec:train-baseline}
In much of the recent literature surrounding PINNs, the training process of the neural network is a combination of stochastic gradient descent-based optimisation (e.g. Adam) followed by a quasi-Newton method (e.g. L-BFGS) until convergence (Raissi \textit{et al.,} \citeyear{Raissi2019}; Sliwinski and Rigas \citeyear{Sliwinski2023}). This process has been largely effective for various fluids-based problems but presents issues with training stability for PINNs using a turbulence model (Patel \textit{et al.,} \citeyear{Patel2024}). In this work, we propose that a negative feedback loop drives the issue, where erroneous mean velocity gradients in the initial predictions produce incorrect eddy viscosity production, which, in turn, causes further issues with the predicted velocity field later in the training process.

\begin{algorithm}[h!] \label{algo:baseline}
    \SetAlgoLined
    \caption{Baseline model - Network training process with adaptive learning rate.}
    
    \textbf{Input:} Network parameters $\theta_0 = (\mathbf{W}, \mathbf{b})$, loss function $\mathcal{L}(\theta)$. \\
    Set equation to RANS-F \\
    \For{$t \gets 1$ to $100{,}000$}{
        Optimise $\theta_t \gets \texttt{Adam}(\mathcal{L}(\theta_{t-1}), \gamma)$. \\
        \If{plateau detected over $5{,}000$ steps}{
            $\gamma \gets \gamma \cdot \gamma_{\text{plateau}}$
        }
    }
    Set equation to RANS-SA. \\
    $\gamma \gets 10^{-4}$ \\
    \For{$t \gets 1$ to $50{,}000$}{
        Optimise $\theta_t \gets \texttt{Adam}(\mathcal{L}(\theta_{t-1}), \gamma)$. \\
        \If{plateau detected over $5{,}000$ steps}{
            $\gamma \gets \gamma \cdot \gamma_{\text{plateau}}$ 
        }
    }
    $\gamma \gets 1$ \\
    \For{$t \gets 1$ to convergence}{
        Optimise $\theta_t \gets \texttt{L-BFGS}(\mathcal{L}(\theta_{t-1}), \gamma)$
    }
    
    \textbf{Output:} Final network parameters $\theta_t$. \\
    
    \textbf{Defaults:} $\gamma = 10^{-3}, \gamma_{\text{plateau}} = 0.5, \theta_0 = (\mathbf{W} \sim U\big(-\sqrt{3}/2, \sqrt{3}/2 \big), \mathbf{b} \gets 0)$.
\end{algorithm}

To overcome this issue, we propose splitting the training procedure into two distinct sections. Firstly, the PINN undergoes training using the RANS-F equation set with just an \textit{Adam} optimiser to stabilise the velocity field to a more physical regime. Following this, a second \textit{Adam} optimisation step is taken at a reduced learning rate to mitigate any significant changes in the velocity field. The training procedure concludes using a \textit{L-BFGS} optimiser until convergence. Algorithm~\ref{algo:baseline} provides the full details, and this procedure will apply to all \textit{baseline} models in Section~\ref{ch:results}.

\begin{table}[h!]
\tabcolsep=0pt%
{\color{black}
\TBL{\caption{Summary of dimensions of the inner data box, driving quantities and their sampling frequency, and driven quantities for model validation of the baseline models\label{tab:baseline-data}}}
{\begin{fntable}
\begin{tabular*}{\textwidth}{@{\extracolsep{\fill}}l cccc ccc@{}}\toprule%
\multicolumn{1}{}{} & \multicolumn{4}{c}{\TCH{Data box}}& \multicolumn{3}{c}{\TCH{Quantities}}
 \\\cmidrule{2-5}\cmidrule{6-8}%
\TCH{Model} & \TCH{$x_{min}$} & \TCH{$x_{max}$} & \TCH{$y_{min}$} & \TCH{$y_{max}$} & \TCH{Driving} & \TCH{Points per chord} & \TCH{Driven} \\\midrule
$L10_b$ & -5 & 10 & -5 & 5 & $\overline{U}, \overline{V}$& 20 & $C_p$, $\overline{u'v'}$ \\
$P75_b$ & -0.1 & 1.3 & -0.6 & 0.5 & $\overline{U}, \overline{V}$& 20 & $C_p$, $\overline{u'v'}$ \\
$R10_b$ & -5 & 10 & -5 & 5 & $\overline{U}, \overline{V}, \tilde{\nu}$ & n/a & $C_p$\\
$R75_b$ & -5 & 10 & -5 & 5 & $\overline{U}, \overline{V}, \tilde{\nu}$ & n/a & $C_p$ \\
\botrule
\end{tabular*}%
\end{fntable}}
}
\end{table}

Alongside giving us a reference of the performance of a PINN with different datasets of the same problem, the baseline models provide a starting point for applying transfer learning. \textcolor{black}{Table~\ref{tab:baseline-data} provides a summary of the four models that will be trained using Algorithm~\ref{algo:baseline} that will act as a source of comparison to the transfer learning cases ($L10_b$ \& $P75_b$) and as an initialisation for the transfer learning models ($R10_b$ \& $R75_b$).}

\subsection{Transfer learning} \label{sec:train-tl}
\textcolor{black}{One of the shortcomings of PINNs is their inherent embedding of the flow configuration in the network, limiting direct application to a new problem, such as a change in geometry, without additional steps or modification to the network architecture. However, for a sufficiently similar flow problem, it is reasonable to assume that the resulting solution, or network parameters for a PINN, would be comparable. Therefore,} this limitation can also be seen as a feature, as many problems rely on a fixed geometry but vary parameters such as the Reynolds number. One approach to leverage this feature is through transfer learning. 

Transfer learning is the process in neural networks of fine-tuning a pre-trained model on a new dataset to enhance its predictive capabilities for a specific task related to that dataset (Weiss \textit{et al.,} \citeyear{Weiss2016}). This technique commonly appears in fields such as large language models (LLMs) and computer vision tasks, where training one of these models from scratch would be prohibitively expensive or yield worse overall results. In the context of PINNs, we can take a pre-existing model and re-train the model at a reduced learning rate with a new dataset, replacing the existing data loss term within the loss function. 

\begin{algorithm} \label{algo:tl}
    \SetAlgoLined
    \caption{Transfer learning training process - RANS-SA only.}
    \textbf{Input:} Network parameters $\theta_0$, loss function $\mathcal{L}(\theta)$. \\
    \For{$t \gets 1$ to $10{,}000$}{
        Optimise $\theta_t \gets \texttt{Adam}(\mathcal{L}(\theta_{t-1}), \gamma)$
    }
    $\gamma \gets 1$ \\
    \For{$t \gets 1$ to convergence}{
        Optimise $\theta_t \gets \texttt{L-BFGS}(\mathcal{L}(\theta_{t-1}), \gamma)$
    }

    \textbf{Output:} Final network parameters $\theta_t$. \\
    \textbf{Defaults:} $\gamma = 10^{-4}, \theta_0 = (\mathbf{W}, \mathbf{b})_{\text{baseline}}$
\end{algorithm}

There are many permutations of transfer learning available with varying levels of freedom given to the new network to update its parameters through the number of iterations, learning rate or available parameters to update (otherwise referred to as frozen/unfrozen layers) (Torrey and Shavlik \citeyear{Torrey2010}). As outlined in Algorithm~\ref{algo:tl}, the process for the work presented here is quite simplistic, reducing only the number of iterations and learning rate. This decision helps reduce the parameter space to search when generating these models, but further optimisation of this process through more complex procedures may be possible. For all models using transfer learning through Section~\ref{ch:results}, Algorithm~\ref{algo:tl} has been used to train the new model.

\begin{figure}[h!]
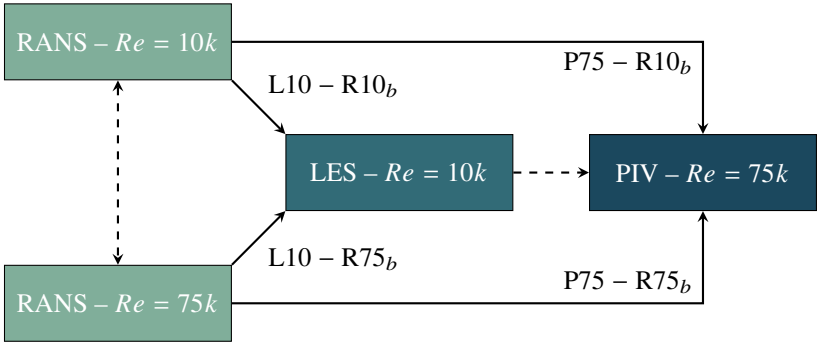

    \centering
    \input flowchart.tex
    \caption{Flowchart of possible transfer learning options with solid arrows indicating those investigated (with associated model name labelled - subscript "$b$" indicating baseline model) and dashed arrows indicating those disregarded}
    \label{fig:TL-flow}
\end{figure}

This method proposes applying a high-fidelity dataset to a single baseline model of a matching geometry. To investigate this, as shown in Figure~\ref{fig:TL-flow}, we generate baseline models from a low-fidelity RANS dataset for two Reynolds numbers and then apply transfer learning to two high-fidelity datasets from LES and PIV for equivalent Reynolds numbers. \textcolor{black}{For the four models in Table~\ref{tab:tl-data}, we consider matching Reynolds numbers and the feasibility of scaling up or down from the baseline model Reynolds numbers. This study omits the application of transfer learning between similar fidelity data, as we are only interested in transitioning from low to high fidelity.} Table~\ref{tab:tl-data} also summarises the sampling frequency and quantities provided for training and validation, respectively.

\begin{table}[h!]
\tabcolsep=0pt%
{\color{black}
\TBL{\caption{Summary of dimensions of the inner data box, driving quantities and their sampling frequency, and driven quantities for model validation of the transfer learning cases\label{tab:tl-data}}}
{\begin{fntable}
\begin{tabular*}{\textwidth}{@{\extracolsep{\fill}}l cccc ccc@{}}\toprule%
\multicolumn{1}{}{} & \multicolumn{4}{c}{\TCH{Data box}}& \multicolumn{3}{c}{\TCH{Quantities}}
 \\\cmidrule{2-5}\cmidrule{6-8}%
\TCH{Model} & \TCH{$x_{min}$} & \TCH{$x_{max}$} & \TCH{$y_{min}$} & \TCH{$y_{max}$} & \TCH{Driving} & \TCH{Points per chord} & \TCH{Driven} \\\midrule
$L10-R10_b$ & \multirow{2}{*}{-0.5} & \multirow{2}{*}{1.5} & \multirow{2}{*}{-0.5} & \multirow{2}{*}{0.5} & \multirow{2}{*}{$\overline{U}, \overline{V}$}& \multirow{2}{*}{20} & \multirow{2}{*}{$C_p$, $\overline{u'v'}$} \\
$L10-R75_b$ & & & & & & & \\\cmidrule{2-8}
$P75-R10_b$ & \multirow{2}{*}{-0.1} & \multirow{2}{*}{1.3} & \multirow{2}{*}{-0.6} & \multirow{2}{*}{0.5} & \multirow{2}{*}{$\overline{U}, \overline{V}$}& \multirow{2}{*}{20} & \multirow{2}{*}{$C_p$, $\overline{u'v'}$} \\
$P75-R75_b$ & & & & & & & \\
\botrule
\end{tabular*}%
\end{fntable}}
}
\end{table}

\begin{table}[h!]
    \centering
    \tabcolsep=10pt%
    {\color{black}
    \TBL{\caption{Summary of PINN training algorithms and computational time \textit{box}\label{tab:train-alg-summary}}}
    {\begin{fntable}
        \begin{tabular*}{\textwidth}{lccc}
            \toprule
            \TCH{Algorithm} & \TCH{Adam iterations} & \TCH{L-BFGS iterations} & \TCH{Wall time (hours)} \\ \midrule
            \TCH{1. Baseline}    & \multicolumn{1}{c}{150,000} & \multicolumn{1}{c}{50,000 or convergence} & 10\\
            \TCH{2. Transfer learning} & \multicolumn{1}{c}{10,000} & \multicolumn{1}{c}{50,000 or convergence} & 1 (+10 for baseline) \\ \botrule
        \end{tabular*}
    \end{fntable}}
    }
\end{table}

One advantage of transfer learning is the subsequent reduction in training time after the first model. For the baseline models trained using Algorithm~\ref{algo:baseline}, the total training time is approximately 10 hours using an NVIDIA A100 GPU, whereas those using the transfer learning process only require one hour on equivalent hardware. \textcolor{black}{The full details of training time in terms of iterations and wall time is summarised in Table~\ref{tab:train-alg-summary}.} This cost-saving highlights one of the other potential advantages, particularly when applying large datasets to PINNs.

%% file: flowchart.tex
\definecolor{t-green}{RGB}{128,174,154}
\definecolor{t-blue}{RGB}{50,107,119}
\definecolor{t-dblue}{RGB}{27,72,94}

\tikzstyle{rans} = [rectangle, minimum width=3cm, minimum height=1cm, text centered, draw=black, fill=t-green, text=white]
\tikzstyle{les} = [rectangle, minimum width=3cm, minimum height=1cm, text centered, draw=black, fill=t-blue, text=white]
\tikzstyle{piv} = [rectangle, minimum width=3cm, minimum height=1cm, text centered, draw=black, fill=t-dblue, text=white]
\tikzstyle{arrow} = [thick,->,>=stealth]
\tikzstyle{arrow2} = [thick, dashed, ->, >=stealth]
\tikzstyle{arrow3} = [thick, dashed, <->, >=stealth]

\begin{tikzpicture}[node distance=1cm, auto]
    \node (l10) [les] {LES -- $Re=10k$};
    \node (r10) [rans, above left =of l10] {RANS -- $Re=10k$};
    \node (r75) [rans, below left =of l10] {RANS -- $Re=75k$};
    \node (p75) [piv, right =of l10] {PIV -- $Re=75k$};
    \draw [arrow] (r10.south east) -- node[anchor=south west]{L$10\ -\ $R$10_b$} (l10.north west);
    \draw [arrow] (r10) -| node[anchor=north east]{P$75\ -\ $R$10_b$}(p75);
    \draw [arrow] (r75.north east) -- node[anchor=north west]{L$10\ -\ $R$75_b$} (l10.south west);
    \draw [arrow] (r75) -| node[anchor=south east]{P$75\ -\ $R$75_b$} (p75);
    \draw [arrow2] (l10) -- (p75);
    \draw [arrow3] (r10) -- (r75);
\end{tikzpicture}

%% file: 4-results.tex
\section{Results} \label{ch:results}
\textcolor{black}{The results section follows the order of the models defined in Section~\ref{ch:setup}. Section~\ref{sec:p75b} outlines the difficulties of applying high-fidelity datasets directly to PINNs with the $L10_b$ and $P75_b$ models. Section~\ref{sec:base-models} applies the same baseline training methodology to the RANS datasets for models $R10_b$ and $R75_b$. Section~\ref{sec:les-tl} presents the transfer learning, initially applied to the LES dataset, as a validation case of the proposed methodology, and the importance of hard constraints. Section~\ref{sec:p75-tl} then applies the transfer learning method to the PIV data. Section~\ref{sec:overview-tl} summarises the findings from the results section and presents the results for the transfer learning models where the baseline Reynolds number is different from the high-fidelity dataset.}

\subsection{PINNs with high-fidelity data} \label{sec:p75b}
To highlight some of the issues with working from \textcolor{black}{high-fidelity} data directly with PINNs, we first created \textcolor{black}{two} PINN models without using transfer learning. \textcolor{black}{These models, $L10_b$ and $P75_b$, follow the} creation procedure of the baseline model process outlined in Algorithm~\ref{algo:baseline} and parameters outlined through Section~\ref{ch:setup}. For all comparisons, we will consider their relative difference where the error, $\epsilon$, is given as,

\begin{equation} \label{eq:error}
    \epsilon = \boldsymbol{\hat{X}} - \boldsymbol{X},
\end{equation}

\noindent for the driving variables. For the driven variables, we will only consider the structures and magnitudes of the predicted quantities against the reference values.

\begin{figure}[h!]
    \centering
    \includegraphics[width=\linewidth]{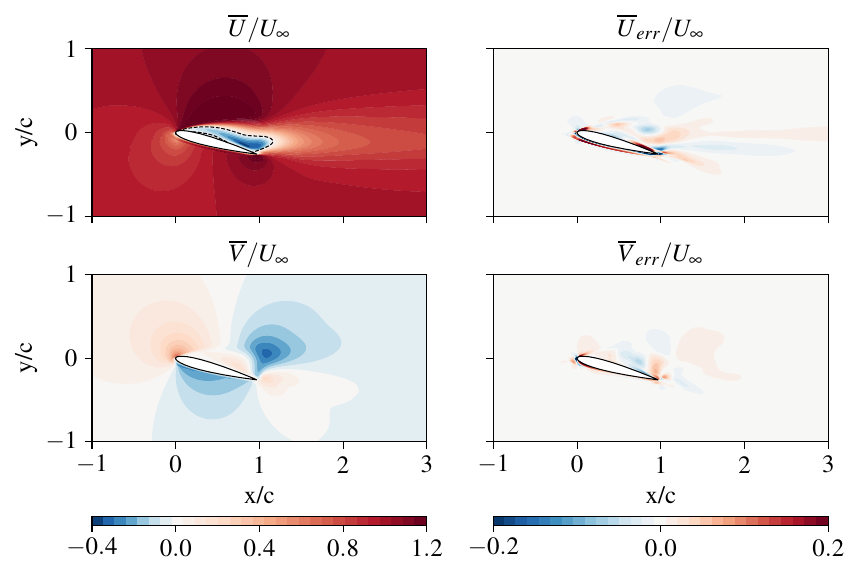}
    \caption{\textcolor{black}{PINN predicted fields from model L$10_b$ of the driving quantities for each velocity component, $\overline{U} \ \& \ \overline{V}$, and their respective errors to reference L10 dataset. For $\overline{U}$, the dashed line indicates the edge of the recirculation region from the experimental data}}
    \label{fig:l10-baseline-UV}
\end{figure}

\textcolor{black}{Figure~\ref{fig:l10-baseline-UV} shows the prediction of the driving variables ($\overline{U} \ \& \ \overline{V}$) and their error, $\epsilon$. As the velocity components are in the data loss for the PINN, it is expected to get a good agreement between the reference fields and the predictions. For the bulk of the wake, there is a small residual error, but the large structures, such as the recirculation region, are closely matched. The largest error is near the airfoil surface on the pressure side of the wing. Whilst the data is sub-sampled, data points are provided up to the surface of the airfoil with no additional masking of the near-wall region. With sparse data, it is reasonable to assume that the PINN is unable to reconstruct the correct velocity gradients, and the sensitivity to those gradients has resulted in the PINN optimising towards a non-physical solution in this region of the domain. Therefore, to improve on this, we would likely need more data close to or on the wall.}

\begin{figure}[h!]
    \centering
    \includegraphics[width=\linewidth]{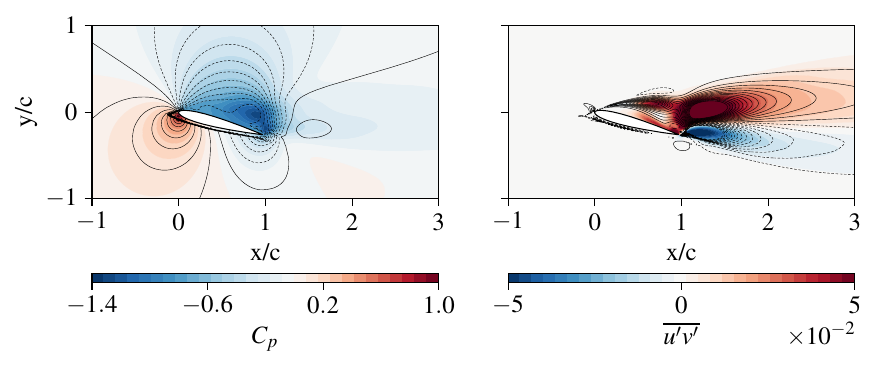}
    \caption{\textcolor{black}{PINN predicted fields (line contours) from model L$10_b$ and reference LES fields (\textcolor{black}{filled} contours) of the driven quantities including the pressure coefficient, $C_p$, and Reynolds shear stress, $\overline{u'v'}$. Solid lines represent positive values, with dashed lines representing negative values for the PINN predictions}}
    \label{fig:l10-baseline-Puv}
\end{figure}

\textcolor{black}{Figure~\ref{fig:l10-baseline-Puv} highlights this non-physical solution near the airfoil's lower surface and how it propagates to neighbouring regions of the domain. Whilst away from the surface of the airfoil, the pressure contours have a good agreement with the reference data; the main issue is focused towards the leading edge. On closer inspection, it appears the surface of the airfoil has been offset slightly upstream, but this is just an unphysical prediction of the pressure. The conclusion is that the PINN tends towards an unphysical solution when there is negative feedback between the velocity gradients and the equations, as the pressure is only constrained by the physics loss here. A similar conclusion can also be drawn when looking at the Reynolds shear stress in Figure~\ref{fig:l10-baseline-Puv}, where the bulk of the prediction away from the airfoil has a reasonable agreement but struggles close to the airfoil. This suggests that with more data around the airfoil, we could get much better results as the physics are largely matched elsewhere in the flow.}

\begin{figure}[h!]
    \centering
    \includegraphics[width=\linewidth]{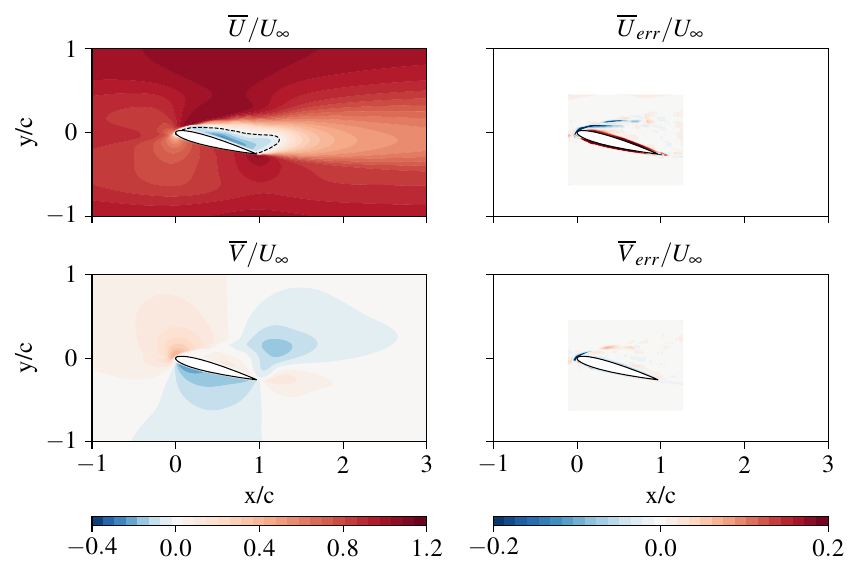}
    \caption{PINN predicted fields from model P$75_b$ of the driving quantities for each velocity component, $\overline{U} \ \& \ \overline{V}$, and their respective errors to reference P75 dataset. For $\overline{U}$, the dashed line indicates the edge of the recirculation region from the experimental data}
    \label{fig:p75-baseline-UV}
\end{figure}

\textcolor{black}{In Figure~\ref{fig:p75-baseline-UV}}, we can see that the presence of a turbulence model allows us to accurately predict the recirculation region with good agreement with the PIV data denoted by the dashed line. Outside of the PIV window, in the far wake region, the contours show that the turbulence model is correctly applying the additional dissipation for the velocity field to recover physically. However, we can see where the PINN struggles using this baseline methodology when considering the error plots. For example, with PIV, resolving the near wall can be challenging due to the presence of reflections on the surface from the laser. Therefore, we could attribute errors in that region to poor reference data rather than the PINN failing to model the physics correctly. In this case, the errors are in three distinct regions: the stagnation point, the near wall and the shear layer.

Starting with the shear layer, we see a large underprediction in the velocity, which appears to be a secondary recirculation region. This observation is unphysical and can be attributed to an issue with the PINN when working with a complex flow with erroneous data points. For the near wall predictions, the error is likely a PIV issue rather than the PINN, given the consistency in the wall-normal distance of the error along the chord. However, this argument is not applicable for the stagnation point. Relative to the rest of the domain, the stagnation point is a small region with a large amount of physics to capture, which presents an issue for the PINN. With limited data points in this region, the PINN can only rely on the physics loss to model the velocity gradients here. Therefore, as seen by the concentration of error and discontinuous contours, the PINN cannot accurately model the stagnation point. 

\begin{figure}[h!]
    \centering
    \includegraphics[width=\linewidth]{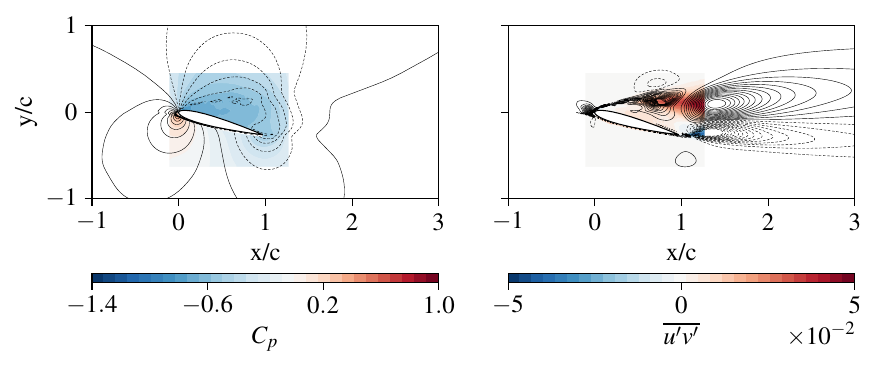}
    \caption{PINN predicted fields (line contours) from model P$75_b$ and reference PIV fields (\textcolor{black}{filled} contours) of the driven quantities including the pressure coefficient, $C_p$, and Reynolds shear stress, $\overline{u'v'}$. \textcolor{black}{Solid lines represent positive values, with dashed lines representing negative values for the PINN predictions}}
    \label{fig:p75-baseline-Puv}
\end{figure}

The driven variables in Figure~\ref{fig:p75-baseline-Puv} reinforce the observations of this model's predictions. From the pressure, we can see that the maximum $C_p$ is not on the airfoil surface. Additionally, the lack of a stagnation point is likely affecting the magnitude of the pressure along the suction side with a reduced acceleration over the leading edge. The comparison of the second-order statistics in the form of the Reynolds shear stress showcases the overfitting present in this PINN model. \textcolor{black}{Critically, when discussing overfitting, there are two aspects to consider: overfitting to the data, where the driving quantity errors are high in locations where data is not provided (as we are only giving a coarse, sub-sampled dataset), and secondly, overfitting to the equations, where the driven, or validation, quantities are not physical. In this case, the former aligns with the aforementioned observations and results depicted in Figure~\ref{fig:p75-baseline-UV}. For the latter, this type of overfitting is shown by the Reynolds stresses in Figure~\ref{fig:p75-baseline-Puv}, where} the general structure bears similarities to the PIV data, its presence before the leading edge and discontinuities in other regions of the flow indicate a lack of physical meaning from the turbulence model and forcing. As our predicted Reynolds stresses are computed using the eddy viscosity and mean velocity gradients, from Equation~\ref{eq:boussinesq}, the error seen in Figure~\ref{fig:p75-baseline-Puv} is directly a consequence of the eddy viscosity having large discontinuities and otherwise unphysical features throughout the domain. From this baseline case, it is clear that an alternative approach is necessary for effectively integrating experimental data with PINNs. 

\subsection{Baseline models} \label{sec:base-models}
To implement transfer learning with \textcolor{black}{high-fidelity} data, we first need to create a baseline model. Here, we intend to create a model that captures some of the fundamental features in the flow, such as the position of the stagnation point and no-slip wall. The fine-tuning training procedure can then modify approximated features, such as the recirculation region, to match the high-fidelity dataset. By building the baseline model from a RANS dataset, we can directly use the eddy viscosity field from the CFD data as a data loss to avoid a trivial solution. The key assumption here is that between the different data fidelities, the similarity of the flow features should result in similar structures of turbulence production, and the PINN will only need to slightly modify the predicted eddy viscosity.





\begin{figure}
    \centering
    \includegraphics[width=\linewidth]{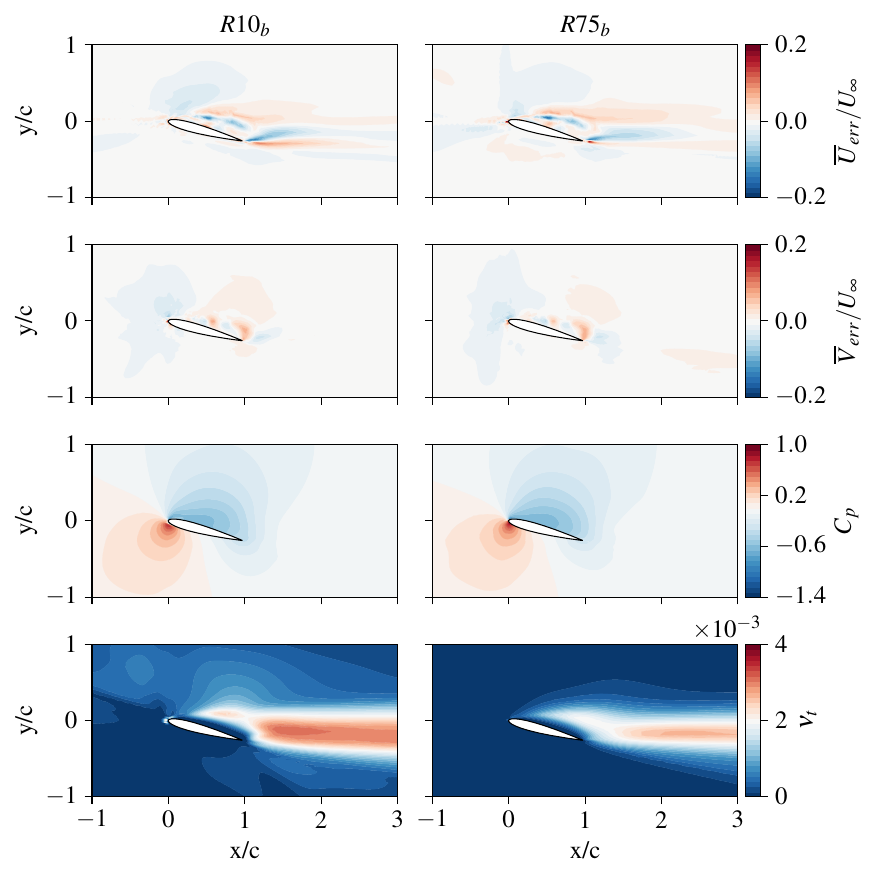}
    \caption{\textcolor{black}{PINN predictions for $R10_b$ and $R75_b$ models for the error in each velocity component, $\overline{U} \ \& \ \overline{V}$, the pressure coefficient, $C_p$, and eddy viscosity, $\nu_t$}}
    \label{fig:r10b-r75b-all}
\end{figure}

\textcolor{black}{Figure~\ref{fig:r10b-r75b-all} shows the predictions of these RANS baseline models.} Across the baseline models, we see a good agreement between the predictions and the reference data outlined in Figure~\ref{fig:ref-r10-r75-all}, with the predicted fields capturing the key structures across all quantities, including those where data was not provided. Unlike the PIV baseline, the RANS baseline models can capture fundamental features such as the stagnation point. The presence of increased near-wall data from the refinement in the CFD mesh likely enables this in the RANS models. This level of resolution and proximity to the wall would be very challenging to acquire through methods such as PIV. Therefore, when utilising these baseline models for transfer learning, the objective is to provide reasonable estimates in regions of the flow for which we would not have data. 

One of the fundamental challenges when implementing a turbulence model is avoiding a trivial solution, given the nature of the transport equation, Equation~\ref{eq:sa-transport}, and the typical magnitudes of $\nu_t$. By providing this quantity in the data loss, we can reduce the likelihood of this occurring, and whilst the accuracy is much closer for the R$75_b$ model, as shown in Figure~\ref{fig:r10b-r75b-all}, some of the unphysical features in the R$10_b$ model highlight that the model is not overfitting to the data. Furthermore, the similarities in both predicted pressures of models R$10_b$ and R$75_b$ reinforce the idea that the PINN trained using this methodology does not overfit and highlight the non-uniqueness in the solutions to the transport equation.

\subsection{PINNs with transfer learning to LES data} \label{sec:les-tl}
One possibility when working with experimental data is that the dataset contains too much noise or otherwise erroneous data points that make applying it to PINNs unfeasible, as the underlying physics of the data may be too far from the truth. For example, the measured velocity field may not obey 2D continuity, which would be true for a stalled airfoil flow that would have some spanwise component. The source of this error could be a combination of experimental uncertainties or missing velocity components in the case of planar PIV. Therefore, to isolate this problem when testing transfer learning, we initially investigated this using an LES dataset. Appendix~\ref{sec:appendixA} contains further information on generating and validating this dataset. For the following sections, all models follow Algorithm~\ref{algo:tl} and use the parameters outlined in Chapter~\ref{ch:setup}.

\subsubsection{\texorpdfstring{\textcolor{black}{$L10\!-\!R10_b$}}{L10-R10b}} \label{sec:l10-r10b}
\begin{figure}[h!]
    \centering
    \includegraphics[width=\linewidth]{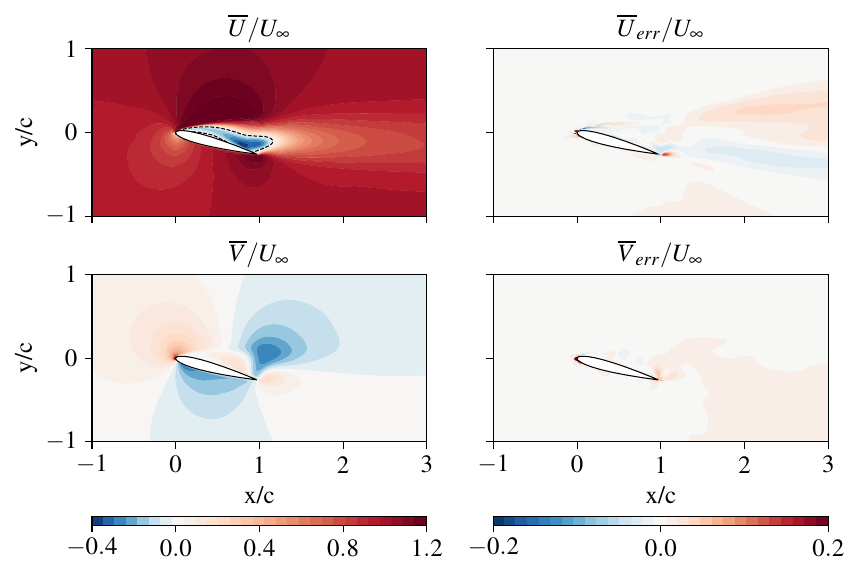}
    \caption{PINN predicted fields from model $L10-R10_b$ of the driving quantities for each velocity component, $\overline{U} \ \& \ \overline{V}$, and their respective errors to reference L10 dataset. For $\overline{U}$, the dashed line indicates the edge of the recirculation region from the LES data}
    \label{fig:l10r10-UV}
\end{figure}

In the first case, we consider the application of transfer learning between datasets of equivalent Reynolds numbers of Re = 10,000. For the PINN, this should be the most straightforward case with both clean data and a lower Reynolds number. From Figure~\ref{fig:l10r10-UV}, we can see that the PINN has been able to modify the velocity field, particularly around the recirculation region. For the streamwise component, the PINN has modified the original recirculation bubble away from the airfoil surface, allowing it to expand in size. The dashed streamline, denoting the reference zero velocity contour, clearly illustrates this ability to capture the new profile. Whilst we cannot directly compare to the PIV baseline (P$75_b$) predictions, the problematic features in that model have either been fully mitigated or reduced in magnitude. This model shows that the error focuses primarily on the leading edge and stagnation point. However, unlike the PIV baseline, this error does not propagate within the shear layer or near-wall regions. 

\begin{figure}[h!]
    \centering
    \includegraphics[width=\linewidth]{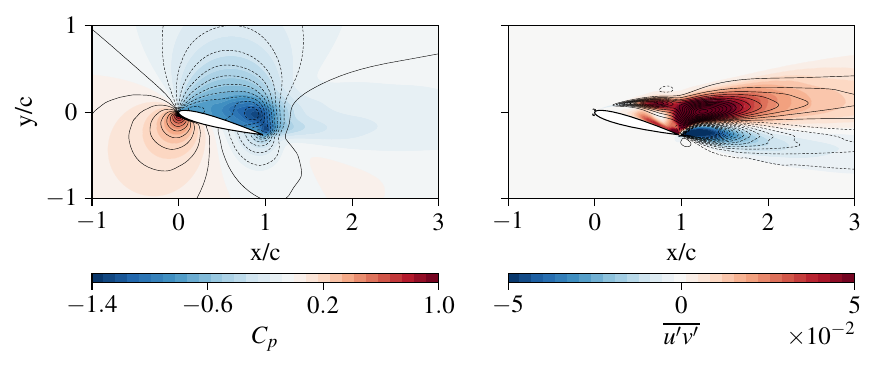}
    \caption{PINN predicted fields (line contours) from model $L10-R10_b$ and LES reference fields (\textcolor{black}{filled} contours) of the driven quantities including the pressure coefficient, $C_p$, and Reynolds shear stress, $\overline{u'v'}$}
    \label{fig:l10r10-Puv}
\end{figure}

We can also examine the driven variables in Figure~\ref{fig:l10r10-Puv} to verify these results and ensure the model does not overfit the new data. For the pressure, we can see that the PINN has recreated some of the defining features, such as the low-pressure region over the trailing edge. However, the limits of the pressure are under-predicted for both the suction side and the stagnation point. Notably, the stagnation point is slightly displaced from the airfoil surface, \textcolor{black}{as indicated by the concentration of error near the leading edge in Figure~\ref{fig:l10r10-UV}.} As with the PIV baseline, this inability to capture the stagnation points likely generates lower pressures downstream.

Regarding the Reynolds shear stress, from Figure~\ref{fig:l10r10-Puv}, we can see that even with a turbulence model, the PINN captures the transition in the shear layer. The exact mechanism that enables the PINN to achieve this is unclear. As the PINN only has access to the sampled mean velocity and the baseline model, this would indicate that the PINN is using a combination of these elements to predict the correct Reynolds stresses. The two sources of information to generate this solution are the sampled mean velocity fields and the baseline PINN model. We can assume that, by definition, the RANS data would not have a transitional shear layer and, therefore, would not be present in the baseline model. Whilst the sampled data is relatively sparse, the gradients evaluated by the collocation points must adjust accordingly to match the values at each location. These velocity gradients, in addition to the wall-normal distance, are responsible for the production of eddy viscosity through the domain. For an equivalent geometry, we would expect some discrepancy, and by only softly enforcing the equations, we get physically consistent results when compared to the LES from a turbulence model not typically used for capturing this transition behaviour. However, more validation of this hypothesis is required. 

\subsubsection{Influence of hard constraints} \label{sec:comp-bcs}

\begin{figure}[h!]
    \centering
    \includegraphics[width=0.9\linewidth]{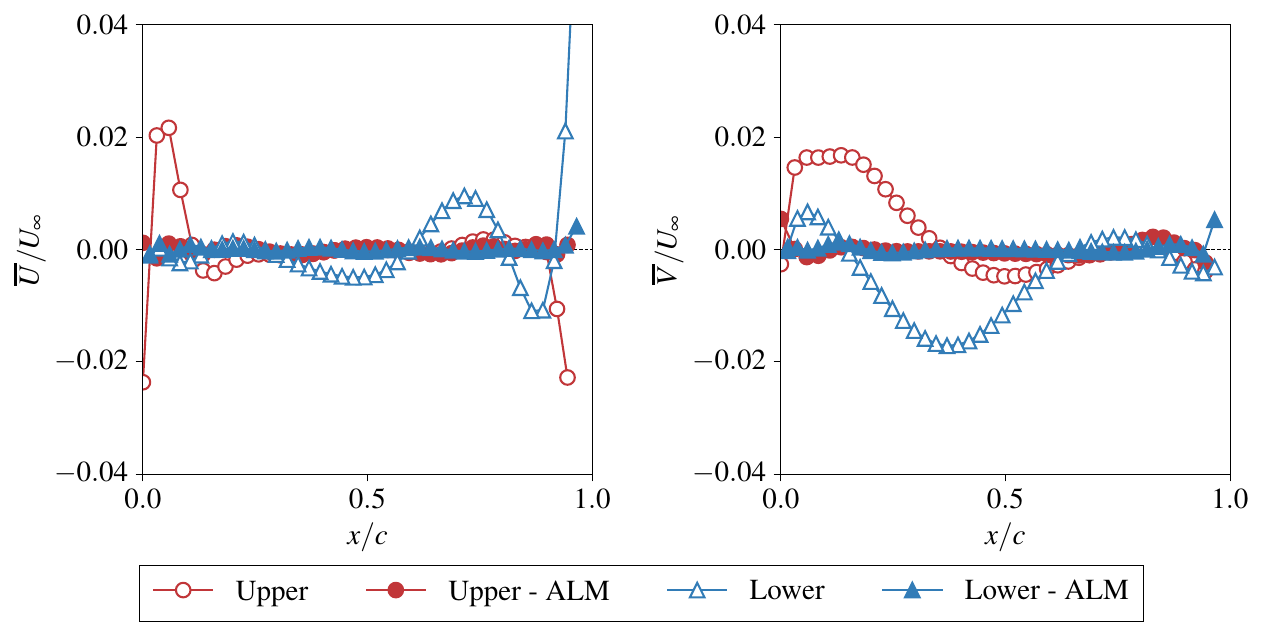}
    \caption{PINN surface predictions comparison between soft and hard (ALM) constraints for each velocity component, $\overline{U} \ \& \ \overline{V}$}
    \label{fig:alm-uv}
\end{figure}

Through the results section, we have not yet considered the quantities at the wall but instead focused mainly on the predictions across the domain. As stated in Section~\ref{sec:hard-bcs}, we have applied a \textit{hard} enforcement of the no-slip wall boundary condition for all models presented. To appreciate why we need to do this and the impact of using hard constraints, Figure~\ref{fig:alm-uv} compares each velocity component along each surface of the airfoil between the L$10\ -\ $R$10_b$ model\textcolor{black}{, from Section~\ref{sec:l10-r10b}, and an equivalent LES model with soft constraints on the wall. Importantly, the baseline model also has soft constraints as the majority of training is done here, as noted in Table~\ref{tab:train-alg-summary}.} Along most of the airfoil, the velocity reduces by as much as two orders of magnitude. For the soft-constrained PINN, we found that the stagnation point is offset upstream of the leading edge, leading to training instabilities and inaccurate predictions, particularly for driven quantities.

\begin{figure}[h!]
    \centering
    \includegraphics[width=\linewidth]{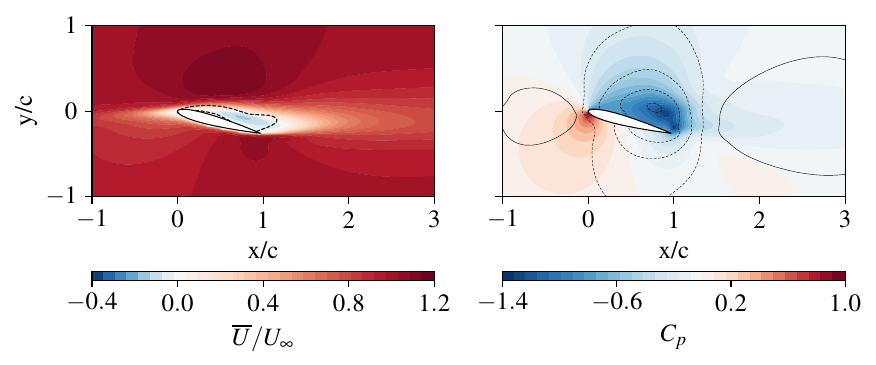}
    \caption{\textcolor{black}{Soft-constrained PINN predictions for streamwise velocity (left), $\overline{U}$, with reference recirculation region denoted by the dashed line, and pressure coefficient (right), $C_p$, with filled contours the reference LES data and line contours the PINN prediction}}
    \label{fig:up-soft}
\end{figure}

\textcolor{black}{To highlight the effects of soft constraints across the entire domain, Figure~\ref{fig:up-soft} shows the predicted streamwise velocity, with the LES recirculation zone denoted by the dashed line, and the LES pressure against the PINN predicted pressure, the line contours. The inclusion of hard constraints not only helps reduce the magnitude of velocity fluctuations but also enforces the airfoil's geometry, capturing its proper curvature and improving pressure predictions.}

\subsubsection{Predicting pressure}

\begin{figure}[h!]
    \centering
    \includegraphics[width=0.65\linewidth]{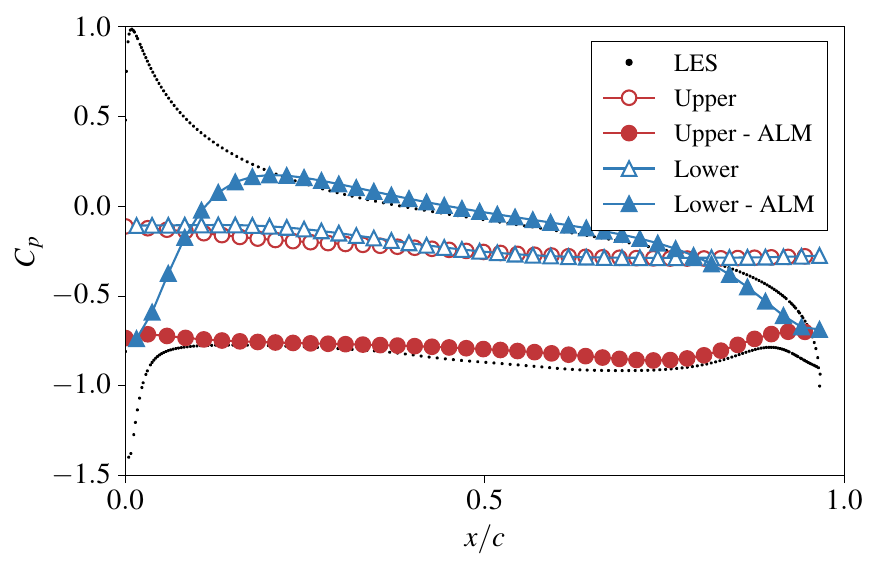}
    \caption{PINN surface pressure coefficient, $C_p$, predictions comparison between soft and hard (ALM) constraints. Reference surface pressure coefficient provided from L10 dataset}
    \label{fig:alm-p}
\end{figure}

Unlike the other quantities on the surface of the airfoil, the pressure acts as the only driven quantity as we do not prescribe any pressure-related boundary conditions. Figure~\ref{fig:alm-p} makes the same comparison as in Section~\ref{sec:comp-bcs} for pressure with the reference surface pressure plotted. Here, we observe that without hard constraints, the pressure remains approximately constant at $C_p \approx -0.2$ around the entire airfoil. Even without the reference data, we can assume this is unphysical for this type of flow and is a product of the velocity fluctuations along the chord, as shown in Figure~\ref{fig:alm-uv}. When we apply hard constraints, after $x/c = 0.2$, the pressure largely recovers to the reference data without any explicit data or boundary conditions enforcing this. This improvement in the driven quantities highlights the importance of correctly enforcing these boundary conditions to get physical predictions. 

As for the points where $x/c < 0.2$, this is likely a combination of some non-zero velocities present on the wall, see Figure~\ref{fig:alm-uv}, and the PINN struggling to model the discontinuity from the airfoil geometry. Whilst no collocation or data points are inside the airfoil, the function generated by the network must be continuous by definition. Therefore, modelling complex physics at the leading edge with small geometries introducing discontinuities makes it very challenging for the PINN to capture these trends accurately. The outcome is that the actual stagnation point is slightly offset from the surface, as seen in Figure~\ref{fig:l10r10-Puv}, which would not appear in Figure~\ref{fig:alm-p}. 

The other aspect of predicting pressure not discussed so far in the results is the coupling of P with $\phi$. To reduce the number of unknowns, we apply the Helmholtz decomposition in Equation~\ref{eq:helmholtz} and group the potential component with the pressure. \textcolor{black}{To recover Equation~\ref{eq:cp-orig} from Equation~\ref{eq:cp}, we must assume} that the $\phi \equiv \phi_\infty$ so that the PINN predicted pressure matches the reference pressure. However, we cannot be certain that this is true as the two components cannot be easily separated. Foures \textit{et al.} \citeyearpar{Foures2014} suggest that the contribution of $\phi$ is small and that $\nabla \phi \approx -\nabla k$ and that ($\overline{P} - \phi$) would be very close to the true pressure, $\overline{P}$. When analysing the other PINN models, in particular where we do not have reliable reference data from the experiments, this method can give us more trust in the predicted pressure fields.

\subsection{PINNs with transfer learning to PIV data} \label{sec:p75-tl}
With the LES models acting as a validation of the methodology, we can now apply the same techniques to the experimental PIV data. We can directly compare the capabilities of transfer learning to the original baseline model in Section~\ref{sec:p75b}. As with applying transfer learning to LES, all models follow Algorithm~\ref{algo:tl} and use the parameters outlined in Section~\ref{ch:setup}. In the following subsections, we reverse the order of the baseline models to start with the case of matched Reynolds numbers before considering variations between datasets.

\subsubsection{\texorpdfstring{\textcolor{black}{$P75\!-\!R75_b$}}{P75-R75b}}

\begin{figure}[h!]
    \centering
    \includegraphics[width=\linewidth]{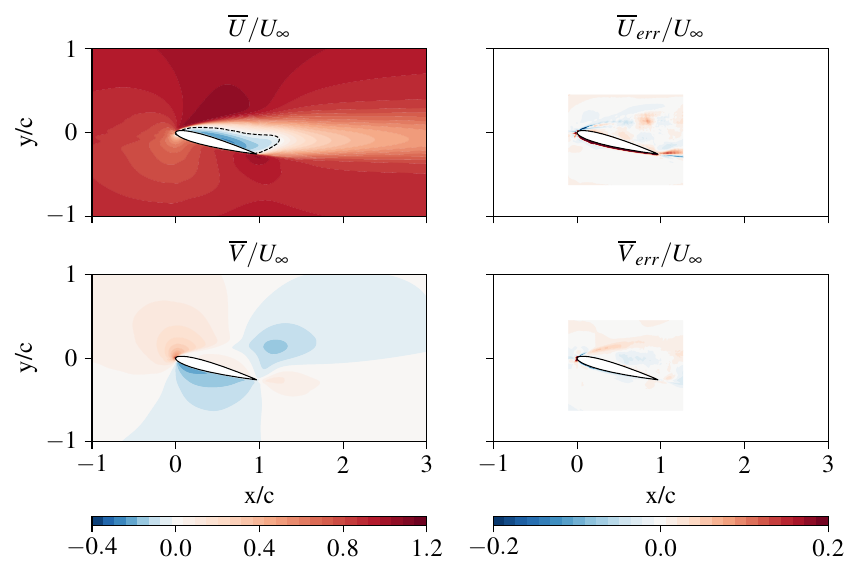}
    \caption{PINN predicted fields from model $P75-R75_b$ of the driving quantities for each velocity component, $\overline{U} \ \& \ \overline{V}$, and their respective errors to reference P75 dataset. For $\overline{U}$, the dashed line indicates the edge of the recirculation region from the experimental data}
    \label{fig:p75r75-UV}
\end{figure}

Figure~\ref{fig:p75r75-UV} outlines the predicted fields for the driving quantities in this PINN setup. From this figure, we can make two comparisons, firstly to the reference data used in training and secondly to the initial baseline attempt without transfer learning. Starting with the reference data, we can see a good agreement for both velocity components. Like the LES transfer learning models, the PINN can accurately modify the recirculation region, as indicated by the zero velocity streamline in the reference data. In this region, we can also see how the PINN can smooth discontinuous data without losing information. These sorts of discontinuities come from post-processing issues of the PIV when multiple cameras or light sources are present. Alongside obtaining otherwise unknown quantities of the flow, PINNs can provide better analysis of experimental data. 

When comparing to the baseline PIV model, we can see that using transfer learning suppresses the presence of unphysical predictions by the PINN. Firstly, we do not get a secondary recirculation region within the shear layer of the leading edge. In the near-wall region, we see a nearly identical error profile on the airfoil's pressure side \textcolor{black}{to the $P75_b$ case. Given the relatively constant distance of the error from the airfoil's surface, it is reasonable to assume that this is from laser reflections and erroneous vectors in this region of the dataset. As the $P75_b$ model did not exhibit any major unphysical features here, it reinforces the point that this error is driven by the reference data. The less complex physics on the pressure side of the airfoil allow the PINN to more easily correct this error over the suction side, which requires additional support.} For the suction side, the error is much lower in the transfer learning case. This indicates that the PINN can more accurately predict the near-wall region for flows with an adverse pressure gradient by providing an initial guess from a baseline model. 

Finally, we can see from the error plots a reduction in the error around the leading edge of the airfoil. The improved shape of the contours at the leading edge suggests a more accurately resolved stagnation point compared to the baseline attempt. The impact of this improvement likely contributes to the aforementioned enhancements to other features of the velocity fields.

\begin{figure}[h!]
    \centering
    \includegraphics[width=\linewidth]{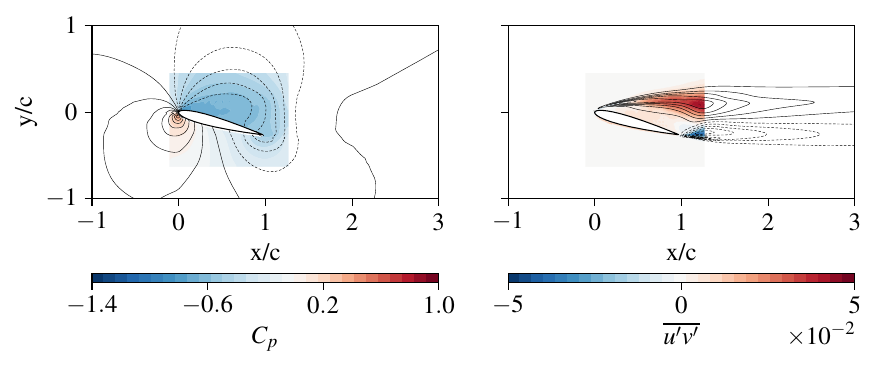}
    \caption{PINN predicted fields (line contours) from model $P75-R75_b$ and reference PIV fields (\textcolor{black}{filled} contours) of the driven quantities including the pressure coefficient, $C_p$, and Reynolds shear stress, $\overline{u'v'}$. Reference $C_p$}
    \label{fig:p75r75-Puv}
\end{figure}

The improvements at the airfoil's leading edge also result in more physical results for the driven quantities. For pressure, we can see the improvements eliminate the detachment of the stagnation point to a position forward of the airfoil. Additionally, in Figure~\ref{fig:p75-baseline-Puv}, some discontinuities can be seen in the low-pressure zone away from the airfoil. In Figure~\ref{fig:p75r75-Puv}, we see a smoother and more continuous variation in the pressure. As for the Reynolds shear stresses, there are significant improvements over the original baseline. The discontinuous and patchy distribution of the stresses through the shear layer are removed and approximate the experimental distribution. We do not want a perfect match for this quantity as the lack of stress in the shear layer from the leading edge is not physical. As previously stated, this is likely an artefact of the experimental data, and the PINN correcting this data indicates that it has learned the underlying physics of the problem, not simply overfitting the model to the data provided. However, the model suppresses the peak magnitude of the stresses when compared to both the baseline and reference data. The three-dimensional nature of the flow is likely to be the source of discrepancy since the PINN is trying to fit the data to two-dimensional equations (Cadambi Padmanaban \textit{et al.,} \citeyear{Cadambi2025}).

\subsection{Performance overview of transfer learning} \label{sec:overview-tl}
\begin{figure}[h!]
    \centering
    \includegraphics[width=0.7\linewidth]{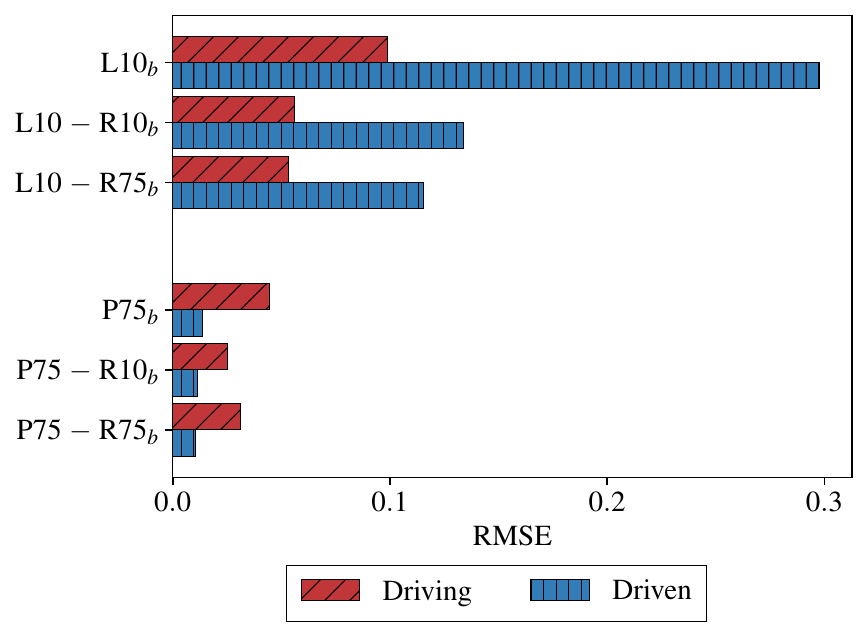}
    \caption{\textcolor{black}{Summary of performance gains when using transfer learning over the baseline algorithm only for driving and driven variables, respectively}}
    \label{fig:tl-summary}
\end{figure}

To supplement these results, Figure~\ref{fig:tl-summary} provides a quantitative overview of the performance boost expected when using transfer learning for a given dataset and baseline model. \textcolor{black}{A direct comparison between these models trained on different datasets is not valid, but all the results are grouped here to show the general trends between the transfer learning cases presented in this work.} For each model outlined in Table~\ref{tab:tl-data} \textcolor{black}{and high-fidelity models in Table~\ref{tab:baseline-data}}, the combined RMSE is computed for both the driving and driven variables, respectively, where the ground truth is available. Notably, this excludes $C_p$ from all P75 models. Therefore, comparisons between the accuracy of models using varying datasets are invalid and are only qualitative in terms of the trends found. The combined RMSE is defined as,

\begin{equation}
    \mathrm{RMSE} = \sqrt{\sum^N_{n=1}\sum^M_{i=1}[\boldsymbol{\hat{X}}_{n,i} - \boldsymbol{X}_{n,i}]^2},
\end{equation}

\noindent where $i$ represents the number of reference points and $n$ is the number of quantities. The RMSE is evaluated with the complete reference data inside the data domain defined in Table~\ref{tab:tl-data}. 

\textcolor{black}{In addition to testing transfer learning with a matched Reynolds number between the RANS and respective high-fidelity dataset, we also considered the effects of applying this methodology with the mismatched Reynolds numbers in the models $L10-R75_b$ and $P75-R10_b$. The proposition for both models is that, given the identical geometries and similarities in the RANS eddy viscosity fields, it may be possible to repurpose a baseline model to avoid significant training time.}

\textcolor{black}{In both models, a small dependence on the quality of the baseline model was apparent, with any errors in the baseline propagating into the final model. This did not cause major differences between the two models for a given dataset, e.g. $L10-R10_b$ and $L10-R75_b$, but any minor discrepancies could only be attributed to the variations in the baseline model. Figure~\ref{fig:tl-summary} reinforces these findings with the RMSE for the driven quantities being lowest for models initialised with the $R75_b$ baseline.} 

\textcolor{black}{For the $P75-R10_b$, the consistency with the $P75-R75_b$ predictions highlights the robustness of the methodology. Additionally, by comparing the error of the two models to the reference PIV data, we could be certain of the errors associated with the PINN and those associated with the uncertainties in the reference data. In particular, those from laser reflections in the near-wall region and the leading edge shear layer.}

\textcolor{black}{In all cases, transfer learning yielded a minimum improvement of 18\% across all variables and as high as 61\% for the driven variables in the $L10-R75_b$ case. The key takeaway from this analysis is that to obtain unknown quantities accurately, the quality of the baseline model supersedes the necessity to match the Reynolds number.}

%% file: 5-conclusion.tex
\section{Conclusion} \label{ch:conclusion}
To summarise the work presented here, \textcolor{black}{we have showcased a new transfer learning-based methodology for improving PINN performance using sparse, high-fidelity measurements. We have demonstrated the robustness of the methodology by first applying it to a LES dataset and then to a higher Reynolds number PIV dataset, with consistent improvements in both cases. These results show that the combination of extensions presented here, transfer learning and hard constrained boundary conditions, helps to overcome the} fundamental challenges within the PINN methodology and enable the recovery of other mean quantities, including pressure and Reynolds shear stress. 

When applying our PIV dataset naively to a PINN model, we found only a reasonable agreement between the driving variables ($\overline{U} \ \& \ \overline{V}$) and the original reference data. From the predicted Reynolds shear stresses, we identified that the source of this error was related to problems predicting eddy viscosity in a physically consistent manner. With transfer learning applied from a RANS-trained PINN baseline model, we observed significant improvements to the prediction of both the driving and the driven quantities. This improvement in the driven quantity predictions indicates that the PINN is learning the underlying physics from the equations and not overfitting the data. Additionally, the ability to derive physical meaning from the quantities learnt indirectly from the model helps improve interpretability, which is of growing interest in the AI and machine learning communities. 

Alongside this new methodology, we also studied the importance of matching all physical constraints on the flow, including the no-slip boundary condition. Here, we used an augmented Lagrangian method to update a Lagrange multiplier on each boundary collocation point to increase the effective weighting of that point to the total loss of the PINN. This extension to the constraints resulted in a reduction of the surface velocity by two orders of magnitude. When comparing the surface pressure, the ALM PINN had a much closer agreement to the reference LES data passed $x/c = 0.2$, where the soft-constrained PINN predicted a near-constant value across the airfoil surface. One possible extension to this would be to investigate the effects of different optimisers and whether more advanced optimisers, such as L-BFGS, can remove deviations, in particular at the leading edge of the airfoil.

Furthermore, we showcased that this methodology is repeatable across different datasets, and transfer learning is applicable between datasets of varying Reynolds numbers. Due to the similarity in the eddy viscosity fields, the transferability between Reynolds numbers indicates that the emphasis should be on creating a higher accuracy baseline model instead of matching the specific Reynolds number of each experimental case in a dataset. This method also helps to reduce the computational cost, particularly over larger datasets. Further exploration is required to verify the extent of this transferability between Reynolds numbers and whether this methodology is appropriate to transfer between similar geometries, for example going from a NACA 0012 airfoil to a NACA 0020. 

%% file: appendix-a.tex
\section{LES data generation and validation}\label{sec:appendixA}

\subsection{Generating dataset}
A high-fidelity computational dataset is generated by performing wall-resolved large eddy simulations (WRLES) of a flow past a NACA 0012 airfoil at \(\alpha = 15^\circ\) using OpenFOAM (Weller \textit{et al.,} \citeyear{weller1998tensorial}). The governing equations are solved with the \texttt{pimpleFoam} solver, which combines the PISO (pressure-implicit with splitting of operators) (Issa \citeyear{issa1986solution}) and SIMPLE (semi-implicit pressure-linked equations) algorithms (Patankar and Spalding \citeyear{patankar1972calculation}). This combination allows for the use of larger time steps, thereby reducing the overall computational cost. The \texttt{pimpleFoam} solver works by seeking a quasi-steady approximation of the solution at each time step, achieved through multiple "inner" iterations. The converged solution at one time step is then propagated to the next time step.  

Turbulence modeling is performed using the wall-adapting local eddy-viscosity (WALE) model (Nicoud and Ducros \citeyear{nicoud1999subgrid}). This model is well-suited for capturing transition and is less dissipative than the traditional Smagorinsky model (Duan and Wang \citeyear{duan2024calibrating}), making it more effective in preserving small-scale turbulence structures. The airfoil has a chord of \(c=1m\) with a blunt trailing edge and is surrounded by an O-grid with a diameter of \(L_x/c = 120\). The pressure and suction sides of the airfoil are discretized using $200$ points each, and the blunt trailing edge is discretized using $15$ points. The domain extends in the spanwise direction by \(L_z/c = 1\), as the flow transitions to a three-dimensional state at this Reynolds number with a dominant spanwise wavelength of \(\lambda_z \approx c/3\) (Gupta \textit{et al.,} \citeyear{gupta2023two}). The span is discretized using $25$ points with a spacing of \(\Delta_z = 0.04c\), giving a total mesh size of $2,266,000$ cells.  

The incoming flow is set at \(U_{\infty} = 1m/s\). The Reynolds number is fixed at \(Re_c = 10,000\) by setting the kinematic viscosity to \(\nu = 10^{-4}m^2/s\). The first grid point is located at \(y_1 = 0.5mm\), resulting in an average \(y^+=0.27\). The gradients are computed using a second-order accurate central differencing scheme. The velocity term is discretized using a limited central differencing scheme that applies an upwind method in regions of strong gradients. The turbulence variable \(\Tilde{\nu}\) is discretized using an unbounded second-order scheme.  The equations for scalar variables are solved using a preconditioned conjugate gradient (PCG) solver, while vector variables are solved using a preconditioned bi-conjugate gradient stabilized (PBiCGStab) solver. A second-order implicit scheme is used for time-stepping. The time step is set to \(\Delta t = 0.001\), resulting in a maximum Courant number of \(Co \approx 2\).  

A mixed boundary condition is applied for velocity and the turbulence variable, switching between a fixed-value Dirichlet type and a fixed-gradient Neumann type based on the sign of the velocity flux. A pressure outlet boundary is imposed along the circumference of the domain, while the airfoil surface is treated as a no-slip boundary. Since \( y^+ < 1 \), the viscous sublayer is fully resolved, eliminating the need for wall functions.  After the initial transient phase, the flow is averaged over 565 eddy turnover times \( T^+ \), where one eddy turnover time is defined as \( T^+ = c/U_\infty \). The integral time scale \( T_i \) is estimated by placing a probe in the wake of the airfoil at approximately \( 0.2c \) downstream of the trailing edge and along the leading-edge shear layer, yielding \( T_i = 0.3 \). Based on this, a sampling frequency of \( f_s = 1/0.3 = 3.33 \) Hz is chosen, resulting in approximately $1800$ instantaneous snapshots of velocity which is then used to compute the time-averaged velocity field. This is span-averaged to obtain the final reference dataset.

\subsection{Validating dataset}
\begin{figure}[!h]
    \centering
    \includegraphics[width=.5\linewidth]{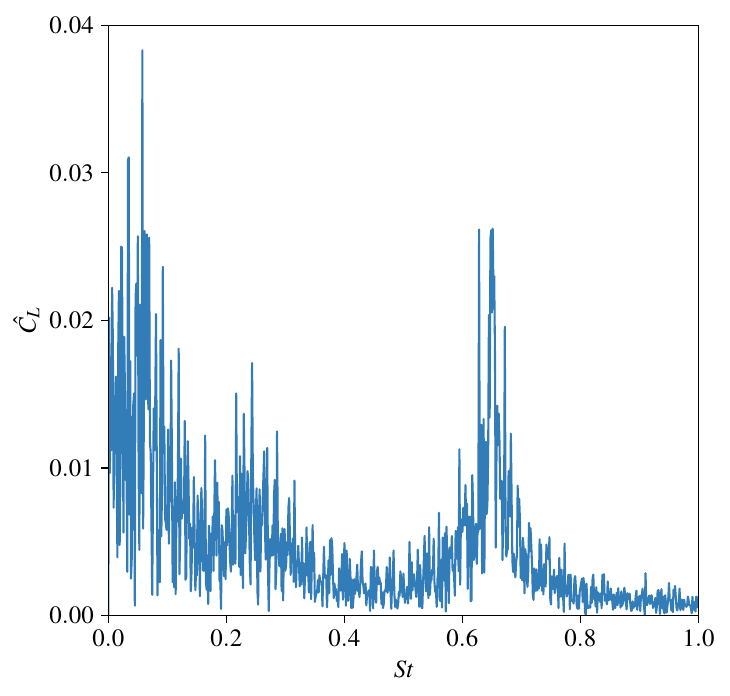}
    \caption{LES lift spectra of a NACA 0012 airfoil at $\alpha$ = 15\textdegree \ and Reynolds number, Re = 10,000}
    \label{fig:lift-spectra}
\end{figure}

While several methods exist to validate the high-fidelity dataset, we present the frequency spectrum of the integrated lift coefficient \( C_L \), defined as  

\begin{align}
    C_L = \frac{L}{\frac{1}{2} \rho U_\infty^2 S c},
\end{align}  

\noindent where \( \rho \) is the fluid density, \( U_\infty \) is the freestream velocity, \( S \) is the span of the wing. The integrated lift coefficient is sampled at a frequency of \( f_s = 10 \) kHz. The corresponding frequency spectrum is shown in Figure~\ref{fig:lift-spectra}. From this, the Strouhal number (\( St \)) is estimated as  

\begin{align}
    St = \frac{f c}{U_\infty},
\end{align}  

\noindent where \( f \) is the dominant frequency.  

The Strouhal number of the peak frequency is \( St = 0.6515 \), which closely agrees with the value reported in Rolandi \textit{et al.} \citeyearpar{rolandi2025biglobal}, where the authors obtained \( St = 0.675 \) for a similar case at \( 14^\circ \) angle of attack and the same Reynolds number. Furthermore, the characteristic three-peak structure in the lift coefficient spectrum, previously observed in Rolandi \textit{et al.} \citeyearpar{rolandi2025biglobal}, is also present in our results.  Additionally, the \( U = 0 \) contour, as shown in Figure~\ref{fig:ref-l10-p75-all}, outlines a small separation bubble slightly downstream of the leading-edge separation point, which is a known characteristic of the flow at this Reynolds number, as also noted in Figure 4 in Rolandi \textit{et al.} \citeyearpar{rolandi2025biglobal}.